# Gemini School
## ASTRONOMY CONTEST WINNER 2010!

**Laser Frequency Comb demonstration with AAT** | **EPOXI mission flyby of UK Schmidt comet**

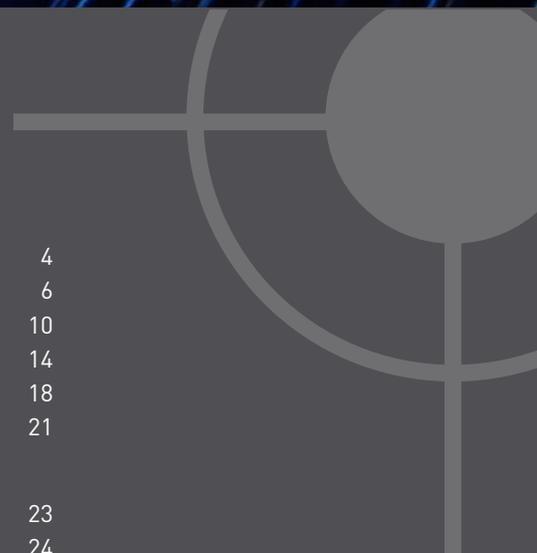

# CONTENTS

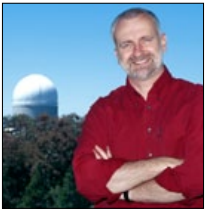

# Director's message

Matthew Colless

This second edition of the new-look Newsletter contains articles drawn from a broad selection of AAO activities. There are reports on the exciting early results from two of the AAT Large Programs – the detection of the baryonic acoustic oscillation signal over a wide range of redshifts by the WiggleZ dark energy survey, and the revealing new data from the GAMA survey on star formation and mass assembly in galaxies. A very different sort of long-term program is described in the article on SPADES, an innovative project combining the talents and resources of professional astronomers and highly professional 'amateurs' to discover and characterise planets orbiting binary stars. There are also reports on two contrasting instrumentation projects – an update from the HERMES team on the design and construction of the major new instrument for the AAT, and an example of the use of the AAT as a testbed facility for new technologies, involving experiments using a laser frequency comb to achieve ultra-precise calibration of spectrographs. Then there are articles describing AAO's support for the cross-fertilisation of scientific ideas by bringing researchers together – en masse for the Southern Cross Conference on *Supernovae and their Host Galaxies* to be held in Sydney in June, and individually via the AAO's Distinguished Visitor Program. Finally, there is a unique first-hand account by Malcolm Hartley of what it's like to discover a comet and then, 25 years later, be present for its spectacular close encounter with a spacecraft!

The major item of news that is not covered elsewhere in the current Newsletter is that the AAO's headquarters in Sydney will likely be moving down Epping Road from Marsfield to North Ryde in 2012. Since it was founded in the mid-1970s, the AAO has occupied land on the Marsfield site owned by CSIRO. Being co-located with the CSIRO Astronomy and Space Science division (CASS) – which operates Australia's radio telescopes and so complements the AAO's role in optical astronomy – has proved valuable for both organisations. However the AAO's current premises are aging, and not gracefully; moreover they are poorly suited to the Observatory's present needs, particularly the requirements of the instrumentation program. The AAO has therefore accepted an offer from the Department of Innovation to provide new premises, to be shared with a branch of the National Measurement Institute. These premises will be constructed at 105 Delhi Road in North Ryde, 5 kilometres down Epping Road towards the city from the current site, and offer a number of benefits besides purpose-built new accommodation. The site is directly adjacent to CSIRO's North Ryde campus and looks over Lane Cove National Park, yet it also offers excellent road and rail access (North Ryde station is just 700 metres away) and a wider range of amenities than the Marsfield site. The planning stage for the new building is now well advanced and construction is expected to start later this month, as soon the project gains final government approval. If all goes to plan, the AAO would move into the new building in the latter half of 2012, staging the transfer to avoid disrupting ongoing projects – in particular, the HERMES team will remain at Marsfield until the instrument is shipped to the telescope. These purpose-built new premises are a clear signal of the strong support provided by the Australian government, and they will undoubtedly help the AAO work more effectively and efficiently in the service of Australian astronomy.

While new headquarters may be the most visible sign of change at the AAO, other changes, and plans for changes, are also taking place. An important aspect of the new governance arrangements for the AAO will be realised in the first meeting of the Australian Astronomical Observatory Advisory Committee, which will be held over two days at Marsfield and Siding Spring on 16-17 March. The AAOAC is the group of eminent representatives of astronomy, academia, industry and government mandated by the *Australian Astronomical Observatory Act* to monitor the status and performance of the AAO and, by maintaining close engagement with the Australian astronomy community, provide strategic advice to the Director and the Secretary of the Department on a wide range of matters relating to the Observatory. The membership of the Committee will be announced shortly, once the members' appointments have been formally approved.

One of the main items on the agenda of the first AAOAC meeting will be the initiation of an *AAO Forward Look* planning process that will develop strategies to deal with the challenges and opportunities of the next five to ten years. The *Forward Look* will build on the national strategies outlined in the Mid-Term Review of the Australian Astronomy Decadal Plan, which is scheduled for release early in March. The issues to be addressed will include: determining the effective scientific lifetimes of the AAT and UKST; optimising the telescopes' productivity for their lifetimes; managing the AAO's changing role at Siding Spring Observatory; improving the AAO's support for off-shore telescopes in which Australia is a partner, such as Gemini and GMT; planning the AAO's next generation of instruments for all these telescopes; exploiting the relocation of the AAO's Sydney headquarters to energise the organisation; and continuing the recruitment and nurturing of world-class staff. The *Forward Look* will involve significant effort by senior AAO staff and consultation with the user community, primarily through the AAO Advisory Committee, the AAO Users' Committee and the Australian Time Allocation Committee, but also more broadly through Astronomy Australia Limited and the National Committee for Astronomy. The *Forward Look* process will run for about six months, with the final report intended to be delivered at the second AAOAC meeting of the year in about September and then made public.

Each of these Newsletters provides a snapshot of just a few of the AAO's activities at one instant in time. In future Newsletters I hope you will be able to read about some of the other projects currently under way or under development. These include the RAVE survey of the Galaxy being conducted on the UK Schmidt Telescope, the proposed GALAH Galactic Archaeology survey using HERMES, the first results from the GNOSIS OH-suppression fibre feed for IRIS2, and the design and capabilities of KOALA, the large new integral field unit for AAOmega.

*Ex AAO semper aliquid novi!*

# Testing the cosmological model using the WiggleZ Dark Energy Survey

Chris Blake (Swinburne University), on behalf of the WiggleZ collaboration

http://wigglez.swin.edu.au

On 13th January 2011 the WiggleZ Dark Energy Survey completed its final (and 245th!) scheduled night of observing. Over the last 4 years we have used the Anglo-Australian Telescope (AAT) to obtain redshifts for over 200,000 galaxies, mapping out the large-scale structure of the Universe over an unprecedented cosmic volume across look-back times up to 8 billion years, over half the age of the Universe. We are now beginning to use this sample to make new and precise tests of the cosmological model. This article describes some of these initial results.

The main goal of the WiggleZ Survey is to measure the properties of the dark energy which appears to constitute the bulk of today's Universe, propelling its expansion into a counter-intuitive phase of acceleration. The apparent existence of dark energy is one of the most perplexing and interesting problems in astronomy today, because it unambiguously indicates new and undiscovered physics. Either Einstein's vision of gravity must be modified on large scales to produce an effective repulsive gravitational force, or the Universe is filled with a new and diffuse source of energy which exerts a negative pressure. The leading suggestion for solving the dark energy problem is that space must be filled with a "cosmological constant", the energy of the quantum vacuum. However, theoretical calculations of this energy density are wildly inconsistent with the amplitude observed by astronomers. This fundamental discrepancy at the heart of physical theory has motivated new and diverse projects to measure the properties of dark energy.

The WiggleZ Survey can probe dark energy using two complementary methods. Firstly, the survey is designed to be large enough to detect the faint imprint of "baryon acoustic oscillations" in the clustering pattern of galaxies. This signature manifests itself as a small preference for pairs of galaxies to be separated by a distance of 150 Mpc. The origin of this effect is extremely well understood as the propagation of sound waves in the hot, dense cosmic plasma in the first 380,000 years of the Universe, prior to "recombination" and the generation of the Cosmic Microwave Background (CMB) radiation.

Measurements of the CMB enable us to calibrate this scale very precisely, and apply it as a standard cosmological ruler in the low-redshift Universe in a manner similar to the use of distant supernovae as standard candles. By mapping out the cosmic expansion history in this geometrical manner, we can measure how the impact of dark energy on the dynamics of the Universe changes with time. And we can do so in a manner complementary to and independent of supernova observations, which must grapple with a number of key systematic effects which modify the "standard candle" luminosity.

The second, simultaneous method through which the WiggleZ Survey can learn about dark energy is to measure how fast the cosmic web of structure grows with time. Clusters and superclusters attract surrounding matter through the force of gravity, setting up coherent "bulk flows" of galaxies through space. The velocities of these flows may be measured because they modify the redshift of each galaxy due to the Doppler effect, creating correlated patterns of redshift offsets between nearby galaxies which are referred to as "redshift-space distortions". The velocity of infall is a direct test of the nature of dark energy because it depends on the force of gravity exerted by each cluster and supercluster. By simultaneously measuring both the growth rate of structure and the overall cosmic expansion, we can distinguish between different physical theories for dark energy.

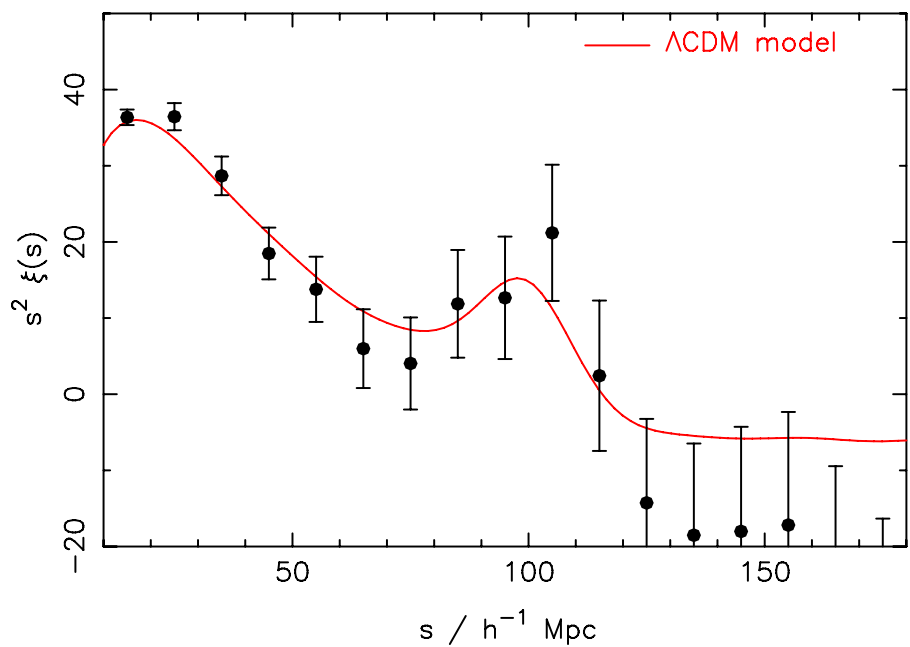

**Figure 1:** Our current detection of the baryon acoustic peak in the clustering pattern of galaxies using 132,000 WiggleZ galaxy redshifts. The final detection will be based on our 200,000 galaxies.

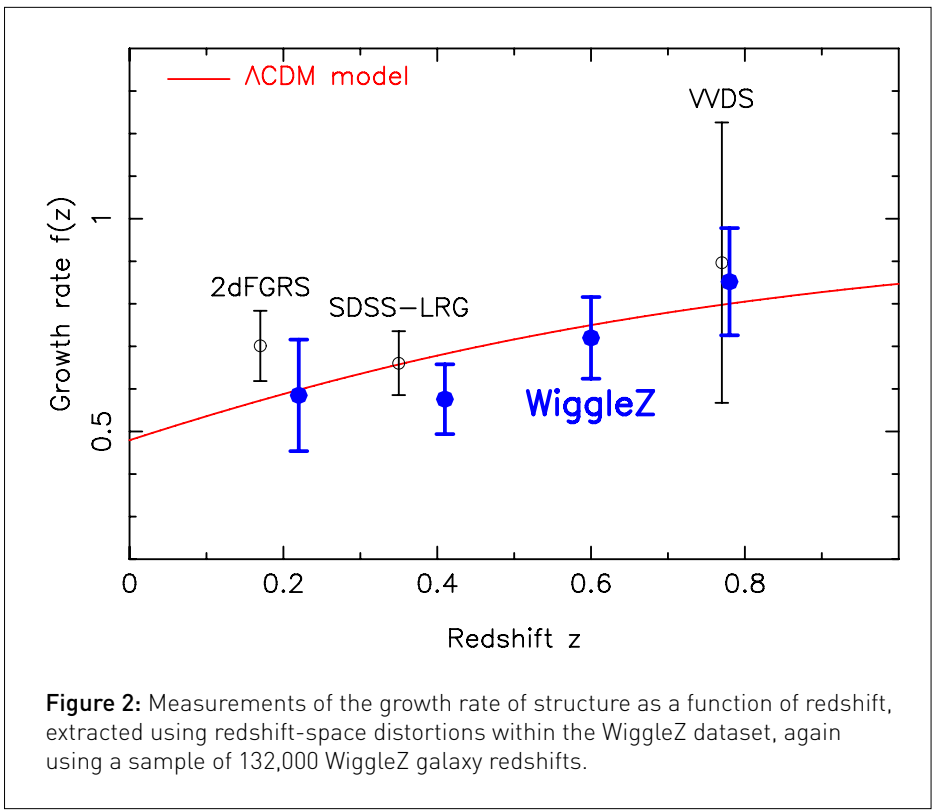

**Figure 2:** Measurements of the growth rate of structure as a function of redshift, extracted using redshift-space distortions within the WiggleZ dataset, again using a sample of 132,000 WiggleZ galaxy redshifts.

We have now produced a preliminary version of each of these measurements in papers to be submitted to Monthly Notices of the Royal Astronomical Society this month. This analysis currently uses 132,000 WiggleZ galaxy redshifts, with final results from the full sample to follow later this year. Figure 1 displays our current detection of the baryon acoustic peak in the clustering pattern of galaxies. This measurement is essentially performed using a simple count of the number of pairs of galaxies with a given separation, which is plotted here as the "galaxy correlation function". The "bump" appearing at a separation 100 Mpc/h (≈150 Mpc) is the expected acoustic peak, which we detect with a statistical significance exceeding 3-sigma. This signature has been previously measured using the Sloan Digital Sky Survey (SDSS) sample at redshift z=0.35; using the WiggleZ data we have extended this measurement to redshift z=0.6, the highest to-date. This subtle imprint in the clustering of galaxies may only be detected with very large surveys mapping hundreds of thousands of galaxies. The solid curve in Figure 1 is the prediction of the standard "Lambda CDM" concordance cosmological model. In this model, the composition of today's Universe is split into approximately 4 per cent baryonic matter, 23 per cent cold dark matter (CDM), and 73 per cent cosmological constant dark energy. This model produces a good fit to the WiggleZ galaxy clustering data, reproducing the observed position of the acoustic peak. Combined with the SDSS measurement of the acoustic peak at lower redshift, we have produced evidence for a cosmological-constant dark energy independently of the supernova observations.

Figure 2 shows the measurements of the growth rate of structure as a function of redshift that may be extracted using redshift-space distortions within the WiggleZ dataset. The open circles indicate the existing measurements prior to WiggleZ. At low redshifts z < 0.4 the growth rate had been well-measured by the 2-degree Field Galaxy Redshift Survey (2dFGRS) and the SDSS Luminous Red Galaxy (LRG) samples. At higher redshifts z > 0.4 there only existed a noisy data point obtained from the VIRMOS-VLT Deep Survey (VVDS) at the Very Large

Telescope (VLT). The measurements from WiggleZ are over-plotted as the solid circles. We have successfully mapped out the growth rate to z=0.8 with a precision comparable to existing low-redshift data. We are the first survey to span such a wide redshift range (and corresponding wide cosmic timeline) within a single project. The solid line is the prediction of the same "concordance cosmological model" as plotted in Figure 1, which also provides a good simultaneous fit to the growth history. Note that as cosmic time advances (redshift decreases) the growth rate slows down, due to the increasing importance of dark energy which acts as a repulsive force to retard the gravitational growth.

We conclude that the cosmological constant model for dark energy provides a good description of the current WiggleZ dataset, which furnishes a test of this model across a much wider range of the history of the Universe than has been previously possible. The agreement between our simultaneous measurements of the cosmic expansion history and growth history implies that General Relativity provides a self-consistent picture of large-scale gravity. The onus returns to theoretical physicists to understand the microscopic nature of how a cosmological constant energy can drive the dynamics of the Universe.

The success of the WiggleZ Survey has rested on the good performance and reliability of the AAOmega spectrograph, which remains one of the best instruments in the world for carrying out large galaxy redshift surveys. We are extremely grateful to AAO staff for their continuing support and development of this facility. The project has also involved a large and significant collaboration within Australian institutions, which have led both the observational programme and the scientific exploitation of the data.

Several further significant results are expected from the WiggleZ Survey dataset during 2011, for example: new limits on the neutrino mass from its influence on large-scale structure; direct measurements of the Hubble parameter at high redshift using geometrical techniques and radial baryon oscillations; tests of inflation using the gaussianity of the density fluctuations; new catalogues of superclusters and voids; and measurements of very large-scale power at the epoch of matter-radiation equality. We anticipate that the WiggleZ dataset will be influential in cosmology for many years to come.

# Galaxy And Mass Assembly (GAMA): Successes, Progress and Plans

Andrew Hopkins (AAO), Sarah Brough (AAO), Ivan Baldry (Liverpool John Moores University), Simon Driver (St. Andrews University), Madusha Gunawardhana (University of Sydney), Jon Loveday (University of Sussex), Aaron Robotham (St. Andrews University), Edward Taylor (University of Sydney), Dinuka Wijesinghe (University of Sydney), and the GAMA collaboration.

http://www.gama-survey.org/

## Introduction

The Galaxy And Mass Assembly (GAMA) survey is a multiwavelength (ultraviolet, optical, NIR, MIR, FIR, radio) photometric and (optical) spectroscopic project aimed at producing the most comprehensive database of galaxies and their properties over the past third of the age of the Universe (Driver et al., 2009). To date GAMA has catalogued over 130000 galaxies spanning 144 square degrees, and aims to expand this to almost 400000 galaxies over 360 square degrees. This survey is poised to address fundamental questions in cosmology related to the nature of gravity and dark energy, along with providing the best measurements of all aspects of galaxy evolution, giving us the most complete picture yet possible of how our Universe works. This article provides an update on the status of the GAMA survey, highlighting some of the exciting scientific results already produced, as well as outlining plans for the coming years.

## GAMA Successes

The GAMA team began observations with the AAT in 2008, with 66 nights of AAT time awarded from 2008-2010. The first public data release of 59000 GAMA spectra occurred on 25 June 2010, announced and demonstrated at the AAO Symposium in Coonabarabran (see AAO Newsletter 118, Aug. 2010). The full sample obtained from these observations encompasses 130000 new spectra of 120000 unique galaxies (Driver et al., 2011), over a survey area of 144 square degrees distributed in three equatorial fields. These observations have been highly successful, with a remarkably high level of completeness, both in target selection and redshift measurement, with the survey overall achieving a 98% redshift completeness to the nominated magnitude limits ($r_{AB}$=19.4 mag in the G09 and G15 fields, and $r_{AB}$=19.8 mag in the G12 field). The redshift distribution of all measured galaxies in the GAMA fields to these limits is shown in Figure 1, and the completeness of the survey is shown in Figure 2.

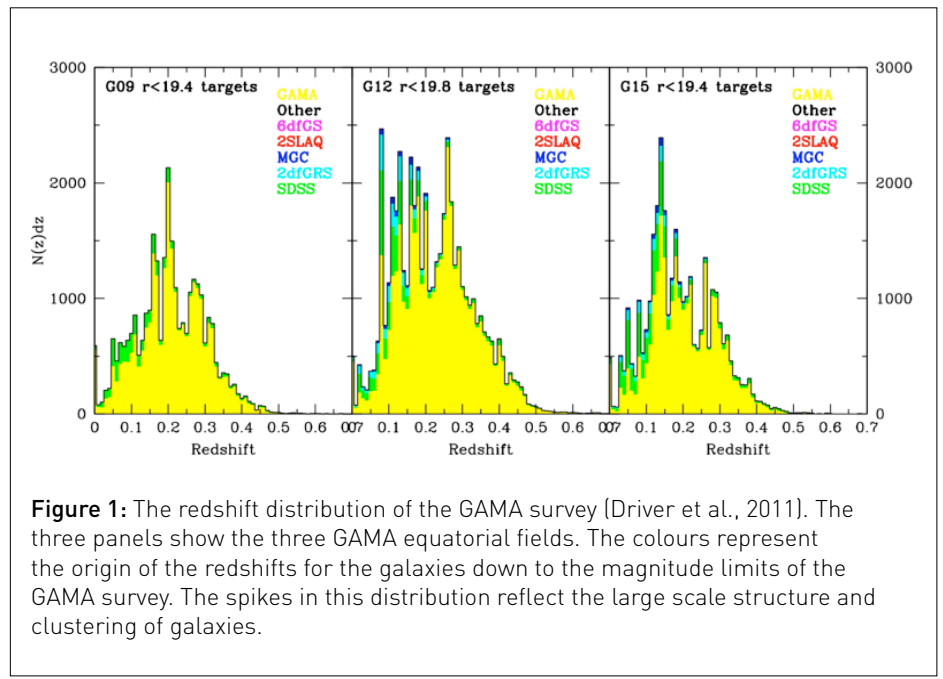

**Figure 1:** The redshift distribution of the GAMA survey (Driver et al., 2011). The three panels show the three GAMA equatorial fields. The colours represent the origin of the redshifts for the galaxies down to the magnitude limits of the GAMA survey. The spikes in this distribution reflect the large scale structure and clustering of galaxies.

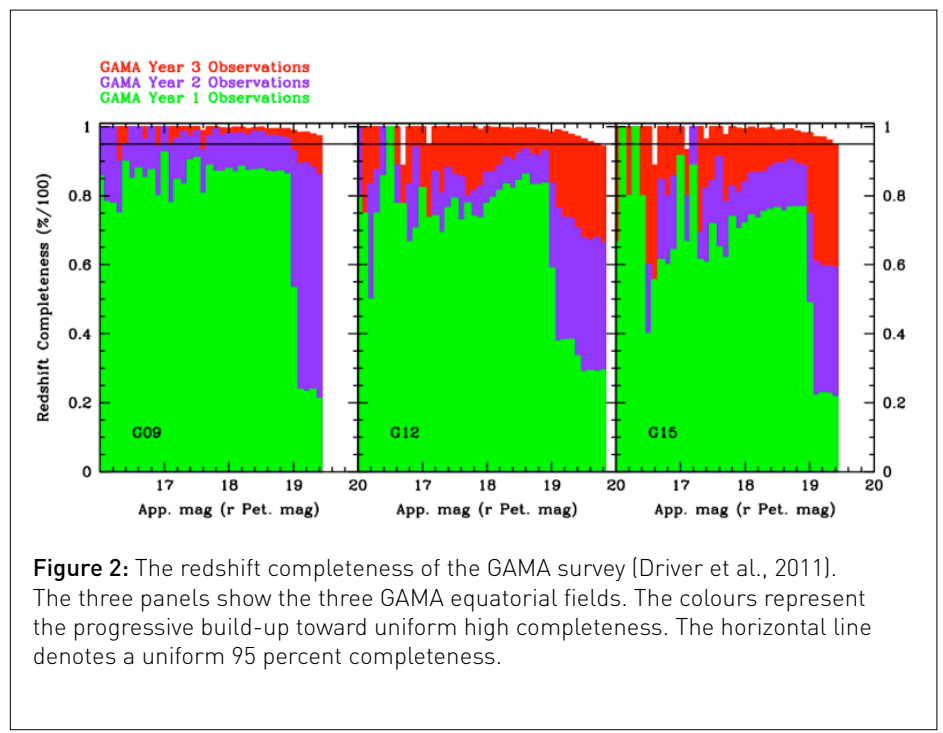

**Figure 2:** The redshift completeness of the GAMA survey (Driver et al., 2011). The three panels show the three GAMA equatorial fields. The colours represent the progressive build-up toward uniform high completeness. The horizontal line denotes a uniform 95 percent completeness.

The scope of the GAMA project, and the strong success of this initial phase, has been a key to facilitating and coordinating with many other surveys that have chosen to focus on the GAMA survey regions. In particular, the *Herschel* ATLAS (PIs: S. Eales, L. Dunne) and the ASKAP DINGO (PI: M. Meyer) surveys are prioritising the GAMA survey regions for their own observations, given the fundamental importance of the optical spectra and redshifts in enabling much of the scientific investigations of those projects. The GALEX medium imaging survey (MIS) has also prioritised the GAMA regions, in a project called GALEX-GAMA (PIs: R. Tuffs, C. Popescu, M. Seibert). Recently, one of the GAMA survey areas has also been chosen as the target for part of the largest XMM project allocated to date, called XXL (10ks to cover 25 square degrees to a depth of ~5x10$^{-15}$ erg/cm$^2$/s).

Detailed analysis of the galaxy properties in GAMA is well underway. A number of the basic properties of the sample are outlined here, taken from a variety of the papers either submitted or in preparation by GAMA team members. The distribution of absolute r-band magnitude with redshift (Driver et al., 2011) showing the GAMA DR1, full sample, and the *Herschel* ATLAS Science Verification data, is shown in Figure 3. The redshift distribution in the G12 field is shown in Figure 4, highlighting the development of the redshift measurements over the course of the survey. Stellar mass measurements for the galaxies in the survey have been calculated (Taylor et al., 2011), and Figure 5 shows the 50%, 80%, 95% and 99% completeness limits in redshift for a given galaxy stellar mass. Figure 6 shows this in a different way, with the redshift limits shown as contours overlaid on the distribution of *g-r* colour vs stellar mass. The distribution of galaxy star formation rates with redshift, and colour-coded by mass, is shown in Figure 7 (Gunawardhana et al., 2011). This illustrates the survey sensitivity to low-mass, low-SFR systems and how this depends on redshift. Measurements of the multiwavelength luminosity functions (Loveday et al., in prep), and the galaxy stellar mass function (Baldry et al., in prep) are also well advanced (see Science Montage pages 16-17).

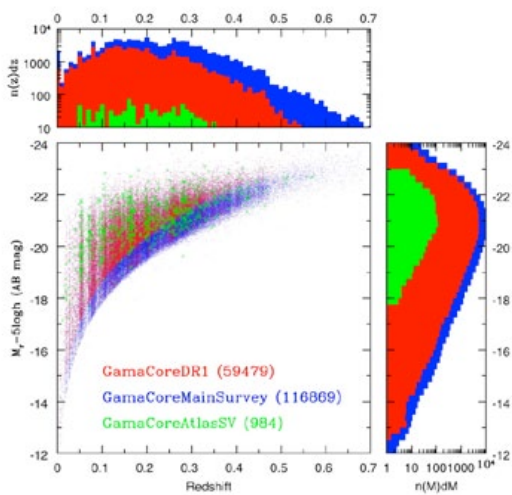

**Figure 3:** The absolute r-band magnitude as a function of redshift for the three science ready catalogues (Driver et al., 2011).

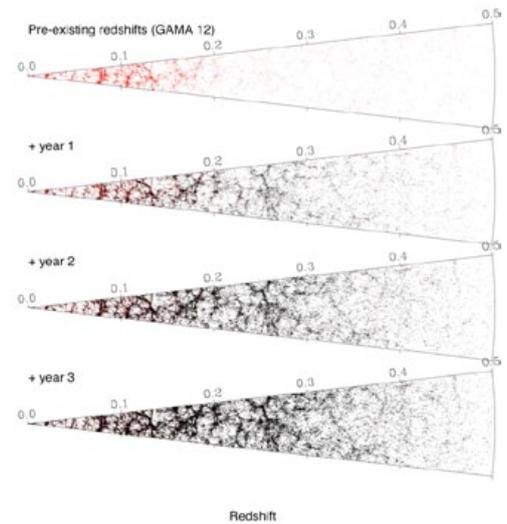

**Figure 4:** The redshift distribution in the G12 field (Driver et al., 2011), showing the progression of the redshift measurements over the initial 3 years of observations.

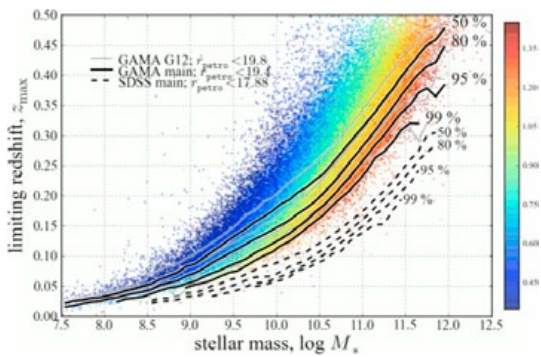

**Figure 5:** The GAMA stellar mass completeness limits as a function of redshift (Taylor et al., 2011). The derived $z_{max}$ (that is, the maximum redshift at which an individual galaxy would satisfy the GAMA r-band selection limit), is used to show the redshift-dependent GAMA mass completeness limits. The stellar masses for which GAMA is 50%/80%/95%/99% complete are calculated in narrow bins of $z_{max}$, with the heavy solid lines showing the r=19.4 main sample limit, and the light solid lines showing the r=19.8 limit of the G12 field. The dashed lines show the equivalent SDSS limits. The data points shown are colour-coded by the rest-frame (g-i) colour.

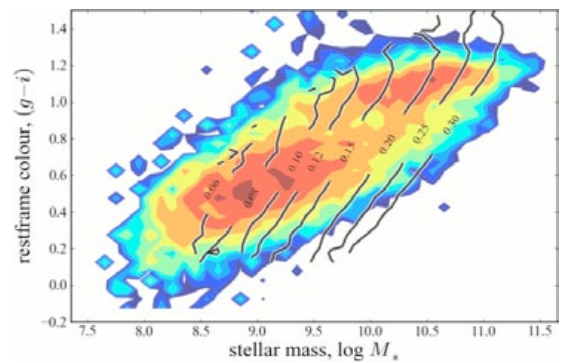

**Figure 6:** The GAMA stellar mass completeness limits as a function of (g-i) rest-frame colour (Taylor et al., 2011). Here the mean $z_{max}$ value is shown as the black contour, overlaid on a data-density colour-contour display of the GAMA colour-stellar mass diagram. The data contributing to these filled colour contours are the incompleteness-corrected bivariate colour-stellar mass distribution of z≤0.12 galaxies. The galaxy red sequence (upper right) and blue cloud (lower left) populations are both visible.

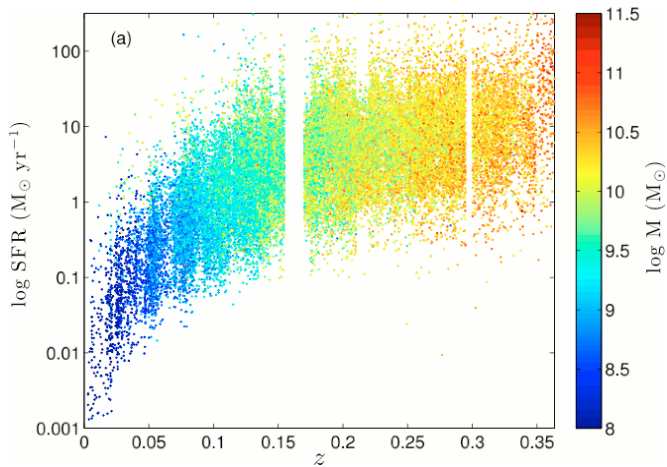

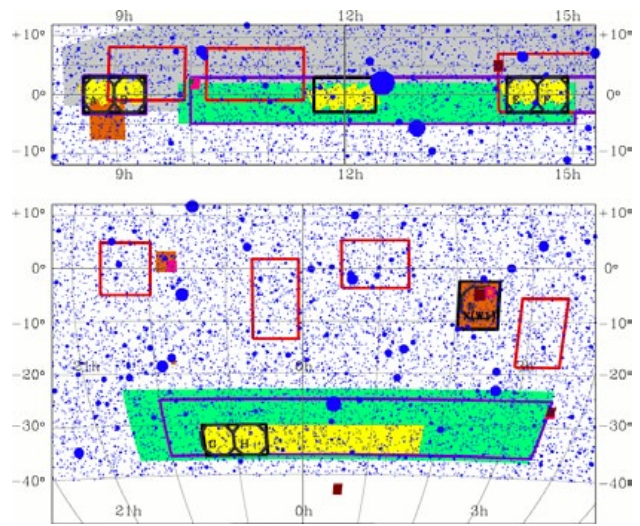

**Figure 7:** The star formation rates of galaxies in GAMA as a function of redshift. The colour-coding is by galaxy stellar mass, and demonstrates the sensitivity of GAMA to extremely low mass and low SFR systems at the lowest redshifts. The redshift bands with no data have been masked out; these correspond to regions where the redshifted Hα line is strongly affected by atmospheric emission features.

**Figure 8:** The GAMA footprint (black rectangles) covering the 3 equatorial survey areas G09, G12, G15 and the two new southern areas G23 and G02. G02 is the CFHTLenS field called W1. Green shows the the 2dFGRS, grey the SDSS, and red rectangles the WiggleZ coverage. Yellow is the *Herschel* ATLAS survey area, and purple rectangles are the VISTA VIKING survey area. The filled blue circles represent NVSS radio continuum sources, with larger circles indicating brighter sources.

## GAMA Progress

Following these early successes, the GAMA project has subsequently been awarded a further 109 nights of AAT time over 2011-2012 to extend the survey area to 360 square degrees, and 375000 galaxies, to address the cosmology and galaxy evolution goals originally identified for the survey. This will be achieved by an increase in the footprint of the equatorial fields, together with the addition of southern fields around 23ʰ and 02ʰ, now referred to as G23 and G02 (Figure 8).

As well as the award of additional observing time, the GAMA team has also been growing. We now have an international collaboration of over 50 members, including 16 students. At the AAO we are fortunate to be welcoming 4 ARC-funded Super Science Fellows in 2010-2011, to contribute to the GAMA survey. Two are already based at the AAO, and two are moving to the AAO in the second half of this year. The AAO will also be hosting a Monash-based Super Science Fellow, for up to one of the three years of that Fellowship. Together with additional growth through new collaboration memberships from new students and occasional postdoctoral appointments, we expect the team to have grown to over 60 members by the end of 2011.

## GAMA Science

The GAMA collaboration has already begun to reap the scientific rewards of the survey. Six full GAMA team papers (Baldry et al., 2010; Robotham et al., 2010; Driver et al., 2011; Hill et al., 2011; Brough et al., 2011; Wijesinghe et al., 2011a), and seven H-ATLAS papers making use of GAMA spectroscopy (Amblard et al., 2010; Dye et al., 2010; Hardcastle et al., 2010; Jarvis et al., 2010; Smith et al., 2011; Guo et al., 2011; Dunne et al., 2011) are currently published or in press. Three GAMA team papers are currently under review (Gunawardhana et al., 2011; Taylor et al., 2011; Wijesinghe et al., 2011b), with six more shortly to be submitted (Baldry et al.; Brough et al.; Foster et al.; Loveday et al.; Robotham et al.; Taylor et al.), and another dozen at an advanced stage and likely to be submitted in the coming year. Including these latter 18, there are over 30 approved paper proposals on the GAMA team wiki.

Some highlights of the scientific results published or in progress are shown in the Science Montage on pages 16-17. A summary of these various outcomes is given here, and the reader is directed to the Science Montage, and to the original references, for further details.

Both galaxy luminosity (Loveday et al., in prep) and mass functions (Baldry et al., in prep) demonstrate the sensitivity of GAMA to the faint galaxy population at low redshift. The luminosity functions show the evolution of both red and blue galaxy populations. Some properties of the least-luminous star forming galaxies contributing to the low-mass systems have begun to be explored (Brough et al., 2011). This demonstrates that these predominantly low-mass, low star formation rate (SFRs less than about 0.01 $M_\odot$yr$^{-1}$) dwarf galaxy systems favour the lowest-density local environments.

Galaxy densities are being quantified through a variety of metrics, and an analysis comparing these with each other is underway to identify potential systematics and preferred approaches depending on the scientific context (Brough et al., in prep). Galaxy groups have been catalogued and quantified by multiplicity (galaxy number), velocity dispersion, halo mass, and related metrics (Robotham et al., in prep). These have been used to identify analogues to our own Local Group, and the full group catalogue now serving, in a complementary fashion to other environmental metrics (Brough et al., in prep) as a reference for environmental-dependence studies within the GAMA survey.

Obscuration properties of star forming galaxies have been explored in detail (Wijesinghe et al., 2011a), demonstrating that an obscuration curve based on models of a turbulent ISM, with the "UV bump" removed, is the most reliable in producing self-consistent SFRs from FUV, NUV, H$\alpha$ and [OII] luminosities. This has been extended by a recent analysis exploring the UV spectral slope $\beta$, and the FIR/FUV luminosity ratio as obscuration metrics for galaxies (Wijesinghe et al., 2011b), which demonstrates that for measurement of galaxy star formation rates, the Balmer decrement remains the most reliable obscuration metric for all SFR-sensitive and obscuration-sensitive luminosities.

Measurement of the stellar initial mass function (IMF) has been traditionally limited to systems where individual stars could be resolved. New approaches are now available that combine stellar population synthesis modelling with integrated galaxy photometry and spectroscopy to allow the inference of the effective slope of the massive end of the stellar IMF for a whole galaxy. Implementing these techniques for the GAMA survey has demonstrated that galaxies with higher SFRs show evidence for flatter (more top-heavy) IMF slopes (Gunawardhana et al., 2011). Further exploration of these results suggests that the underlying dependence is related to the density or compactness of the star formation, with tight relationships between the IMF slope $\alpha$, and the specific SFR or the SFR surface density.

Measuring gas-phase metallicities for galaxies allows exploration of the potential evolution in galaxy heavy-element abundance. The well-known mass-metallicity relationship for galaxies has been measured for many narrow redshift-bin volume-limited samples within GAMA (Foster et al., in prep). Comparison of these results with the local (SDSS-derived) measurement demonstrates that there is little or no evolution in the galaxy mass-metallicity relationship out to z=0.35.

In addition to these developments in our understanding of galaxy properties, the key scientific motivations of the GAMA survey include cosmological measurements that will only achieve the required sensitivity when the redshift survey is complete (Peacock et al., in prep; Norberg et al., in prep). GAMA will play a unique and key part in testing the nature of the cosmic acceleration, distinguishing Dark Energy from modifications of gravity, by spanning the survey area and galaxy mass ranges required in order to measure the mass function of dark-matter halos down to about $10^{12}M_\odot$, more than an order of magnitude smaller than the best current measurements. By measuring the growth rate of structure from the clustering properties of red and blue galaxy populations independently, GAMA will test the precise nature of the theory of gravity, and will set the most robust constraints possible by establishing the level of consistency between these two independent tracers. Preliminary estimates, from the existing GAMA data, have demonstrated that this is feasible with the incorporation of the data anticipated from the 2011-2012 observations.

## GAMA Plans

The public release of complete spectroscopic and multiwavelength photometric data is a primary goal of the GAMA survey. The power of the GAMA survey comes from its complete multiwavelength nature, supplemented by the redshift and spectroscopic information. GAMA will continue an annual cycle of data releases, with the second data release planned for mid-2011, and a third data release encompassing the full 2008-2010 data in early-to-mid 2012. In addition to spectra, images and catalogue information, the public database will incorporate numerous derived and robustly quality-controlled parameters, such as stellar masses, star-formation rates, AGN/SF diagnostic classifications, photometric redshifts, and more, all of which are currently being used by the GAMA team in existing analyses and publications in preparation.

GAMA is not a monolithic edifice. It builds on and ties together numerous collaborations, primarily large photometric surveys including SDSS, UKIDSS, VISTA VIKING, VST KIDS, CFHTLenS, *Herschel* ATLAS, GALEX-GAMA, and ASKAP-DINGO. This coordination of collaborations continues with the development of new projects now building on the GAMA survey itself. Some of these are expected to be led by GAMA team members, others by external collaborators in coordination with GAMA. These include projects to make use of integral field spectrographs (PI: Croom; PI: Brough) that have been successful in obtaining time in the current semester with both the SPIRAL instrument on the AAT and the WIFES instrument on the

ANU 2.3m. Additional plans include the extension of the GALEX-GAMA survey (PI: Tuffs) to prioritise the expanded equatorial and the new southern GAMA fields within the remaining GALEX operational lifespan. Initial discussions with the WISE mid-infrared survey have begun, with an intent to coordinate and collaborate on the incorporation of WISE photometry within the GAMA analyses and plans for deeper spectroscopic follow-up of WISE targets within the GAMA survey regions. And of course, we have been coordinating extensively with the ASKAP-DINGO (PI: Meyer) and ASKAP-WALLABY (PIs: Staveley-Smith & Koribalski) survey projects in anticipation of sensitive neutral hydrogen measurements spanning the full $0<z<0.5$ redshift range of GAMA, when ASKAP begins observations in early 2013.

In summary, the GAMA survey continues to be highly successful, with numerous exciting scientific results in hand and ongoing. GAMA continues to develop new and valuable collaborations and to generate new projects growing from the existing survey. We are looking forward to the exciting results to come from the completion of the AAT observations during 2011-2012. ◆**AAO**◆

# HERMES NEWS

Jeroen Heijmans (AAO), Anthony Heng (AAO), Gayandhi De Silva (AAO), Keith Shortridge (AAO), Lew Waller (AAO), David Orr (AAO) and HERMES team.

http://www.aao.gov.au/AAO/HERMES/

HERMES is the AAO's top-priority project, and has made continuing progress over the last six months, culminating in the Final Design Review (FDR), led by a panel of external experts, just before Christmas. HERMES is a cutting-edge instrument, aiming to produce high-resolution spectra of a large number of target objects simultaneously, and this brings its design close to the limits of the state of the art, particularly with available grating technology. The huge design documentation developed by the team was scrutinized in detail by the panel, The review report provided extensive feedback to the team. While noting some areas of concern, it expressed confidence in AAO's ability to bring the project to a successful conclusion.

Apart from the completion of the detailed design, recent months have seen the lower floor of the AAO's Massey building transformed into the Assembly, Integration and Test (AIT) area for HERMES.  Procurement of parts, some of which have long lead times, has started in earnest.

2011 promises to be an interesting and challenging year for HERMES, as we move ahead into the production phase of the project. A major focus of the FDR was on the recent tests of VPH grating prototypes, which cast doubts on whether the efficiency and mosaicing accuracy required by the conceptual design could be achieved. Further prototypes will be evaluated and fall-back alternatives are being considered. This is the major technical challenge for the project, but the building of an instrument on this scale also presents its share of challenges for project management and systems engineering.

The following sections focus in more detail on various aspects of the project.

## Project management

The momentum generated by the HERMES team's efforts to be ready for FDR will be maintained in 2011. We are now at full steam ahead to complete the detailed mechanical drawings, to complete all the prototype tests, and to procure parts and commence assembly of the components. We still must overcome the major challenge of getting suitable VPH gratings for HERMES, but the mechanical and optical engineers have this as their top priority.

We have already purchased and received four cryostats, CCDs for three of the four channels, various electronic components, the fold mirrors, the fibre optic cable, the slit relay lenses and the fibre prisms. These are all items that require a long test and integration time.

Purchase of all major optical components, except for the gratings, started actually started 6 months before FDR. An FDR for the collimator and the four cameras was carried out successfully in November 2010 with the vendor IRL-KiwiStar. The vendor has now begun manufacture of these components. Coating designs for the camera and the tests to validate the performance are currently being discussed. AAO anticipates receiving the Collimator and three of the four Cameras later this year, with the IR camera delivery planned early 2012.

## Assembly, Integration and Testing (AIT)

An AIT plan has been developed for each of the sub-assemblies. The purpose of these AIT plans is to ensure the team members look at the risks of every facet of the instrument in advance and think of the best way to design, build, and test it.

Testing and integration of the complete instrument is planned for 2012, which would allow transportation to the AAT in early 2013 if no unforseen difficulties arise.

## GALactic Archaeology with HERMES (GALAH)

The primary science driver for HERMES is to undertake a million-star Galactic Archaeology survey. HERMES capabilities are therefore primarily tailored to suit such a survey. The survey team, led by Ken Freeman (ANU) and Joss Bland-Hawthorn (U. Syd), currently consists of over 20 members. In addition to ongoing planning of the survey, the team members have developed a detailed synthetic model abstracts of the Galaxy (Galaxia) to enable target selection and analysis. The team is also at an advanced stage in developing the abundance analysis pipeline (GA3P) to provide the automated abundance determination needed to handle the large volume of survey data. As the FDR board pointed out, the next step is to gain access to the required computing resources.

## HERMES Data Simulator (HDS) and Abundance results

As part of the FDR, there was a presentation of the predicted performance of HERMES, based on a detailed simulation of the instrument as designed.

The HERMES Data Simulator is a command-line interface software tool that generates synthetic detector images of each of the four cameras (spectrograph channels). The HDS system model abstracts the physical design of the HERMES instrument. The goal is that the simulated images are indistinguishable from, or at least closely represent, that expected from the commissioned HERMES instrument. The HDS is an end-to-end instrument data simulator tool providing a mechanism for the early performance and design verification of HERMES. The HDS works with Zemax® to predict the optical performance of HERMES and accepts user-defined data on detector-, star- and sky properties. Details of the simulator can be found in Goodwin et al., 2010, *Proc. SPIE 7735*. An example of the results produced by the HDS is shown in Figure 1 at the top of the page opposite.

The HDS was used to generate 'raw' data for each of the HERMES channels, for various input stellar spectral types, magnitudes, and exposure times.

Using the HDS, a simulated exposure of one hour for a solar type star of V = 14 magnitude yielded a signal-to-noise ratio of 102, 145, 163, 145 per resolution element in the Blue, Green, Red and Infrared channels respectively. While these results are within the required signal-to-noise ratio of 100 per resolution element, the Blue channel clearly has the lowest efficiency and allows for no tolerance on any of the instrument components. One of the recommendations of the review board was to develop an exposure time calculator to keep track of the observation time for the input catalogue.

The simulated output data was then reduced via IRAF and processed via the current version of the GA3P code by Elizabeth Wylie de Boer (ANU). The results show that many of the chemical elements are measureable to within 0.05dex and all other elements are measurable to within 0.1dex accuracy. These results show that science requirements were derived correctly, and also verify that the instrument design meets those requirements.

## HERMES Optics

Optical analyses predict a resolution in normal operating mode ranging from 27,000 to 30,000 depending on fibre and wavelength, which is sufficient to resolve the chemical elements of interest to GALAH. The masked, high resolution, mode results in a resolution range from 47,000 up to 55,000 with a light loss of 47%.

The main concern is the availability of the four large Volume Phase Holographic (VPH) gratings with a record-high blaze angle (67°.2) that are at the heart of the spectrograph. Extensive prototyping with thorough testing at AAO is to continue in the first quarter of 2011. One major issue is efficiency which at currently ~ 50% is on the lower limit of the science requirement for the blue channel. Another is to get an accurate enough mosaic, when the grating is produced by two or more 'exposures' on one substrate.

Following the FDR panel recommendation and with their continuing help, we have put in action a 'tiger' team to look at all aspects, including fall-back options (e.g. mechanically aligned separate gratings), and ensure even closer co-operation with the suppliers.

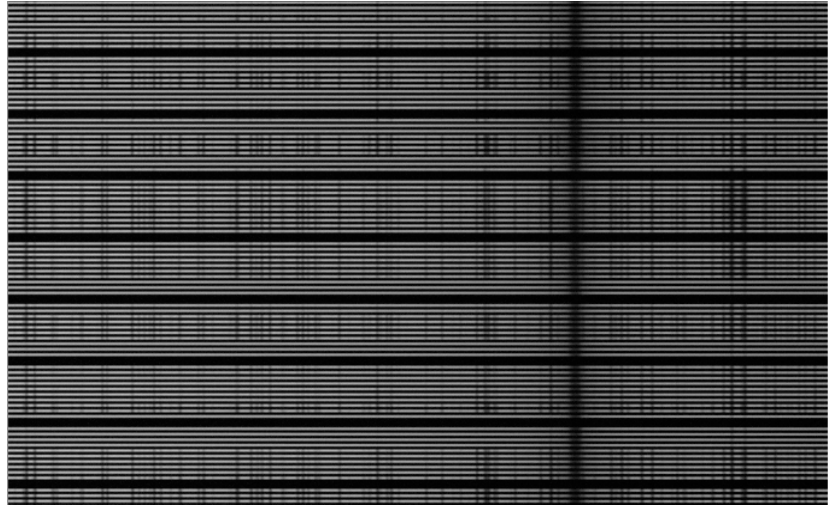

**Figure 1:** Section of the HERMES simulated BLUE channel 470-490nm, for a range of stellar types, clearly showing the H beta line.

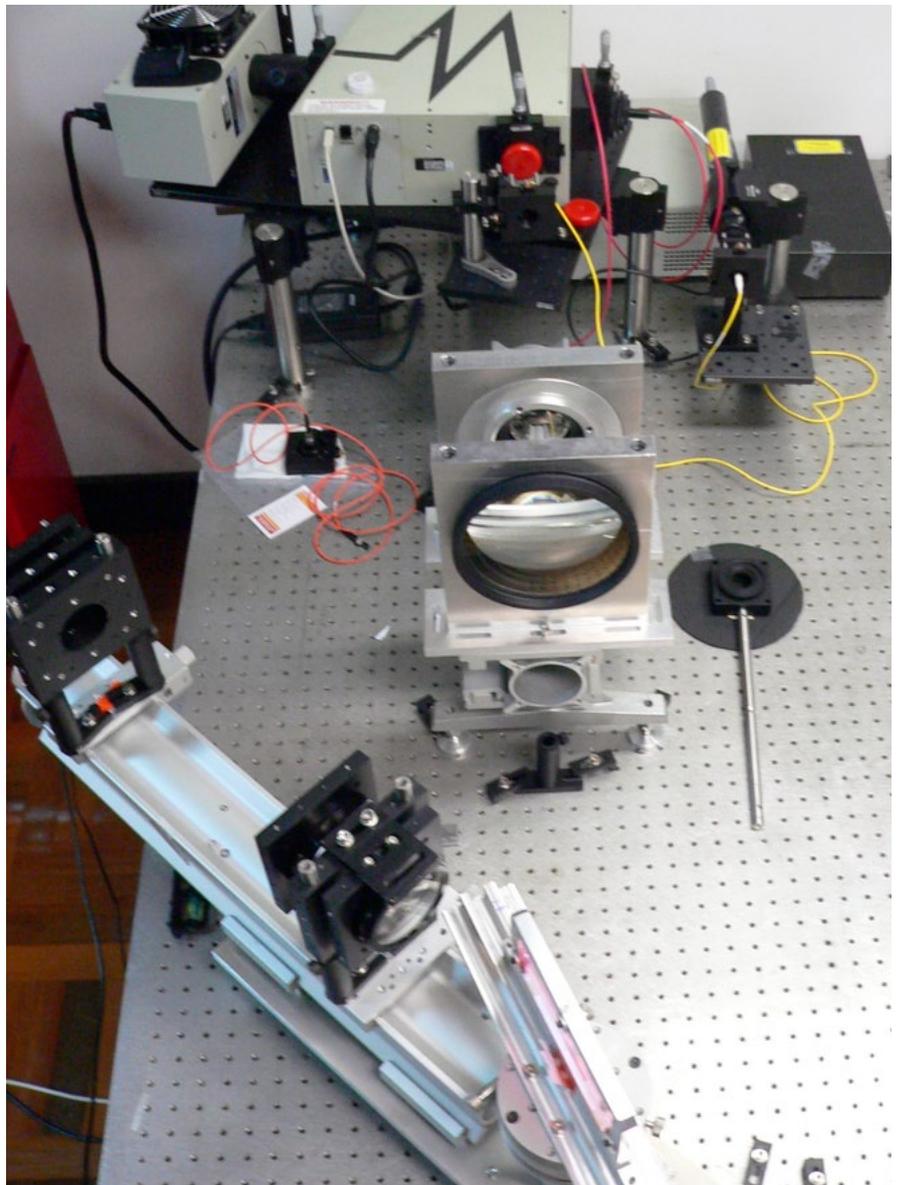

**Figure 2:** Test setup for evaluation of the VPH grating prototypes

## Fibre cable & Slit

The new dual fibre cable design, where AAOmega and HERMES will each get their own fibre cable, was presented at the FDR. The ends of the fibres from the two instruments will come together at the 2dF top end, where each "button" or prism will hold one HERMES fibre and one AAOmega fibre. This requires that a new fibre cable replace the existing AAOmega cable. Test results from fibres with low shrinkage adhesives and retractor modifications show a reduction in focal ratio degradation (FRD). The fibres have been produced and delivered and tests to evaluate a prototype that covers the complete path length are planned for early 2011.

The detailed design of the slit assembly showed how the fibre cable terminates in a V-groove block that holds 40 sets of ten fibres and the relay optics. The 2x40 sets of lenses for the relay optics have already been received. Two pneumatic actuators carry the back illumination modules that can be engaged with the slit when the 2dF field plate is configured.

## Mechanical design

The CAD model of the spectrograph is shown in the figure below. The colours identify the different sub-assemblies. The brown central frame is a partly welded, partly bolted, aluminium structure that supports the complete weight of almost four tonnes on vibration isolators.

Manufacturability and tolerances together with the thermal stability were the main items discussed at the FDR. Finite element analyses supported by test results showed the minimal effect of air temperature fluctuations on the optical stability in the short term (hours).

As the first major optical components will arrive in 2011, extra effort is being put into getting the base frame ready for manufacture. Two additional mechanical designers have been employed from the start of 2011.

## Instrument control

The HERMES instrument control system fits into the existing 2dF/AAOmega control system. The HERMES instrument hardware is controlled via commercial CANbus interfaces. The code to control these CANbus interfaces has been written and tested in simulation, and a prototype CANbus servo system has been assembled and successfully tested. The control system is well advanced and is based on the experience gained from previous instruments at the AAO.

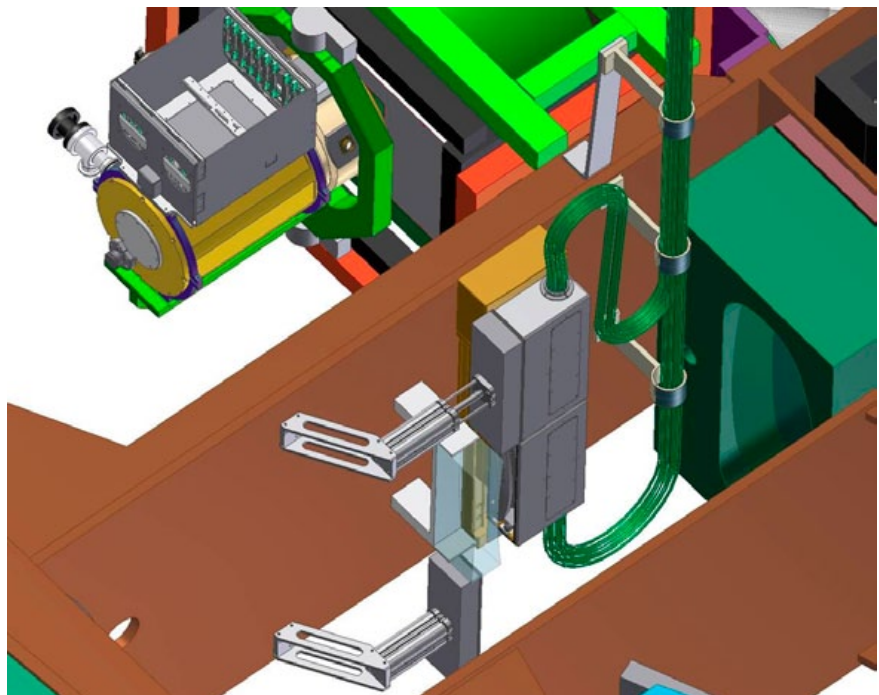

**Figure 3:** Slit assembly with fibre cable (green) mounted in the spectrograph

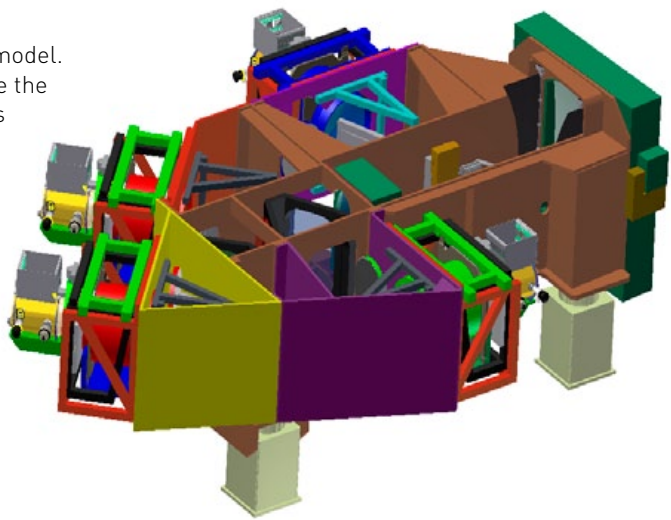

**Figure 4:** The HERMES CAD model. Colours indicate the sub-assemblies

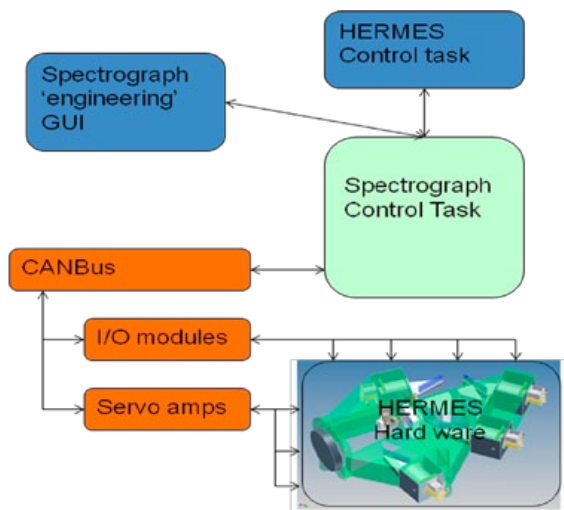

**Figure 5:** Schematic overview of the control architecture

## Detector controller and cryostat

The test cryostat is now cooling the detector below -100°C. Efforts are underway to fine-tune the cooling to achieve the desired operating temperature for the detector.

A prototype of the detector controller electronics to control and read out the e2v CCD231-84 detector has been built and is undergoing tests. The detector controller can read out the detector via one, two or four outputs. Various readout speeds can be selected to trade off read noise against readout time.

An engineering-grade detector and three of the four science-grade detectors have been received from e2v. The red detector is a deep-depletion device with fringe suppression and the IR channel detector will probably be the same type or a bulk-silicon variant. The selection of the IR channel detector type will be made in early 2011.

## Data Reduction Software:

Extensive improvements have been made to the existing AAOmega data reduction system (2dFdr) to enable its use with HERMES data. This includes completion of the optimal fibre extraction code and the implementation of a new wavelength calibration algorithm (now under test).

## Integration room and AAT Instrument room,

The Massey building in Epping now houses the HERMES integration room and an optics and fibre laboratory. The director and the review board officially opened these facilities during the FDR.

The integration room will be used to fully integrate and test HERMES before it is transported to the AAT.

An overview of the designed HERMES room for the AAT is shown below. The footprint of HERMES is some 3.5 x 4.5m. HERMES is accessible via a ceiling mounted crane. Electronic controllers and heat sources are placed

## Science with HERMES workshop

A workshop focusing on the scientific use of HERMES was held on 28-29th September 2010 in Epping.

The purpose of this meeting was to inform the Astronomy community of the capabilities and status of HERMES, and to discuss science possibilities and instrument upgrades. The meeting was a success with over 50 participants, including seven from overseas. The two-day meeting included 10 talks on the various other science projects that can be pursued with HERMES in addition to the planned million-star Galactic archaeology survey. At the conclusion of the meeting it was strongly recommended that the ability to use both HERMES and AAOmega simultaneously, utilising the dual-fibre buttons designed for HERMES, should be implemented.

All workshop presentations are available online at: http://www.aao.gov.au/HERMES/ScienceWorkshop/hermessite_workshop_program

At the end of workshop, the GALAH team had its first meeting. Several working groups were formed which include: the Abundance analysis pipeline; Survey strategy, which will receive feedback from individual working groups on Chemodynamics and chemical tagging, Stellar physics, Stellar reddening and binarity, and synergy with GAIA; the Input catalogue and Observations; Data storage and Release; and Commissioning projects. Participation is open to the Australian astronomy community, as well as overseas astronomers with particular expertise. ✦AAO◦

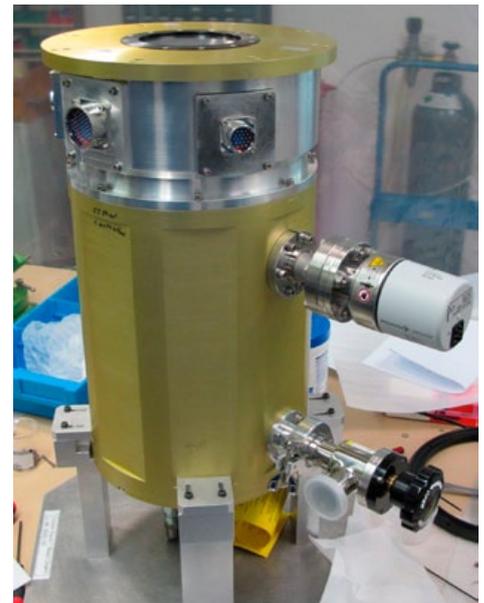

**Figure 6:** Fully assembled test cryostat

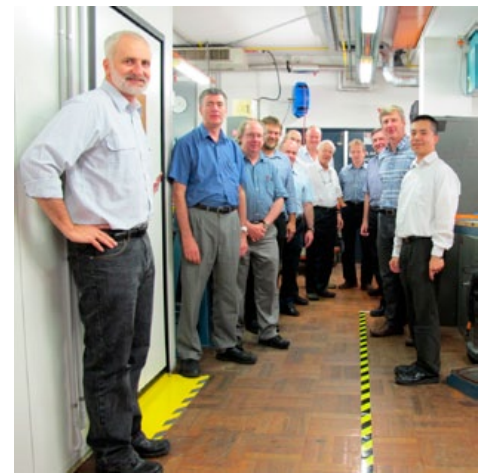

**Figure 7:** Cutting the ribbon of the brand new HERMES integration room in Epping

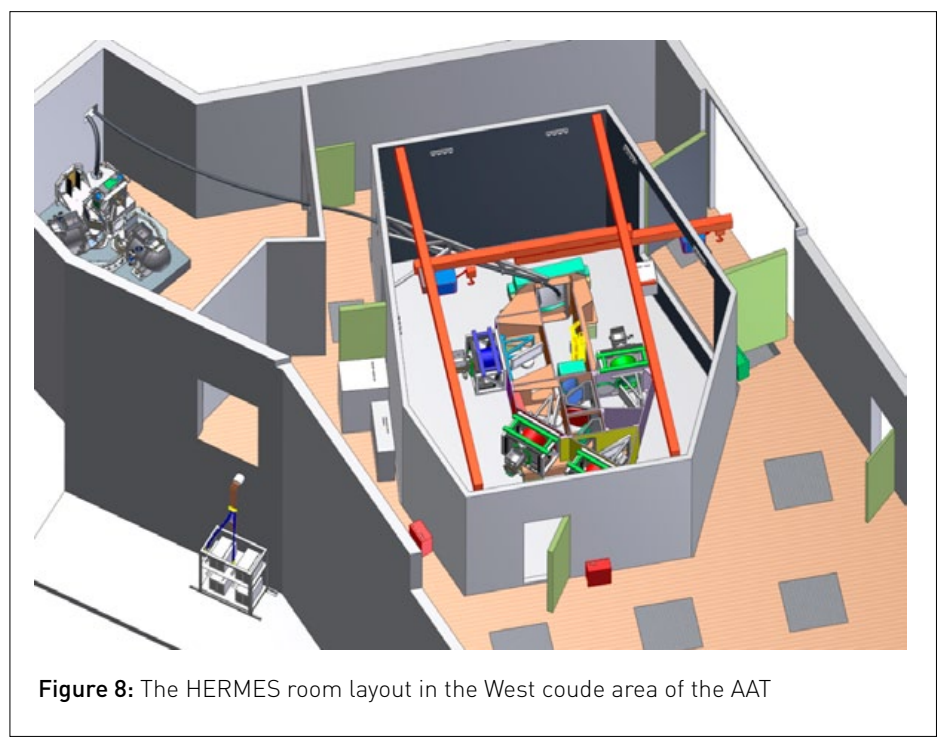

**Figure 8:** The HERMES room layout in the West coude area of the AAT

# Laser frequency comb meets UHRF

Michael T. Murphy (Swinburne University), Clayton R. Locke (University of Western Australia), Philip S. Light (University of Western Australia), Andre N. Luiten (University of Western Australia), Jon S. Lawrence (AAO, Macquarie University), Stuart I. Barnes (AAO)

Having revolutionized metrology – the science of measurement – and earning their creators the 2005 Nobel Prize in Physics (e.g. Hall 2006; Hänsch 2006), laser frequency combs (LFCs) hold significant promise for ultra-precise wavelength calibration of astronomical spectrographs. They provide a regular grid of lines whose absolute frequencies are known a *priori* and to accuracies well beyond the $10^{-11}$ level (~1 cm s$^{-1}$) interesting to astronomers. In April 2010 we tested a LFC on the Ultra-High Resolution Facility (UHRF) at the AAT with the goal of demonstrating the characterization of intra-pixel sensitivity variations in the CCD, something any future "ultra-precise" measurement must achieve.

Many observations require radial velocity accuracy that challenges modern spectrographs and their wavelength calibration. The best-known example – exoplanet discovery – was made possible by employing iodine gas absorption cells or simultaneously-observed thorium emission lamps. Iodine cells only absorb between ~500 and ~650 nm and lamp emission lines are sparse, have huge intensity dynamic range and change subtly with lamp operating conditions and age. This hampers other measurements, for example the search for variations in the fundamental constants using quasar absorption spectra (e.g. Murphy et al. 2007a). One future experiment proposed for the European Extremely Large Telescope (E-ELT) is to measure the change in the Hubble flow with time using the Lyman-alpha forest in high-redshift quasar spectra (Liske et al. 2008). But the change in drift rate is extremely small, just ~1–2 cm s$^{-1}$ yr$^{-1}$: one cannot afford to lose ~50% of photons in an iodine cell and no thorium lamp is stable over the ~30 years required to complete the experiment.

Laser frequency combs (LFCs) may be the answer (Murphy et al. 2007b). Mode-locked, femtosecond-pulsed lasers produce a comb of laser modes in frequency space, $\nu_n = \nu_{ce} + n \, \nu_r$, with the mode spacing determined by the repetition frequency, $\nu_r$, set by the size of the laser cavity. Since $\nu_r$ is in the radio frequency range, it is easily controlled (or just monitored) with simple electronics and can be linked directly with a radio reference, or even an atomic clock for very precise work. The absolute frequency of the n$^{th}$ comb mode, $\nu_n$, then relies on the "carrier offset" frequency $\nu_{ce}$, which can be determined and, importantly, stabilized, by "self-referencing" the comb, i.e. by mixing two different harmonics of the comb and detecting the beat note(s). That is, a self-calibrated LFC provides a series of regular and closely spaced modes, the absolute frequencies of which are known absolutely, and a *priori*, to an accuracy limited only by that of the radio reference. A simple, cheap Global Positioning System receiver provides an absolute comb accuracy of ~$10^{-11}$–$10^{-10}$ or ~1 cm s$^{-1}$.

The first use of an LFC on an astronomical spectrograph (Steinmetz et al. 2008) already demonstrated state-of-the-art calibration precision in the infrared (9 m s$^{-1}$ at 1.5 μm) and an optical test using ESO's vacuum spectrograph, HARPS, achieved calibration repeatability of ~15 cm s$^{-1}$ over several hours (Wilken et al. 2010). Raw precision and repeatability, though, are not enough: measuring anything astrophysical with cm s$^{-1}$ precision will require systematic errors to be characterized in unprecedented detail. However, it should be possible to use LFCs themselves to discover and characterize the instrumental systematics. For example, Wilken et al. easily detected a 512-pixel periodic artifact from the CCD manufacturing process. A subtler problem is intra-pixel sensitivity variations (IPSVs). Individual CCD pixels can show large differences in sensitivity to photons on different parts of their surface (e.g. Toyozumi & Ashley 2005). Pixels can vary in size/shape and may contain small defects. Electron counts may also be registered in a pixel when a photon enters near the edge of a neighbouring one, depending on incidence angle. These, and many other effects, lead to apparent IPSVs. And while the development of deep depletion CCDs has mitigated many of these effects, the behaviour of individual chips – even individual pixels – must be accurately *known*, not just assumed, for truly "ultra-precise" spectroscopy.

With AAO's kind agreement and assistance, in April 2010 we tested an LFC on the Ultra High Resolution Facility (UHRF) on the AAT to demonstrate ways of accurately measuring possible IPSVs. Figure 1 illustrates our set-up. One important 'limitation' with current LFCs is that their fundamental repetition frequencies are too low for astronomical spectrographs to resolve the comb modes from each other. A typical LFC has $\nu_r = 0.1$ GHz but, even in UHRF's highest resolution mode, $R = 940$k, its FWHM resolution is 0.4 GHz. A simple solution employed in previous works is to filter out most modes with a Fabry-Perot cavity. We chose the cavity's free spectral range (FSR) to transmit only every 11th comb mode, leaving an effective mode separation of 1 GHz – wide enough to resolve with UHRF. Interestingly, bypassing the filter cavity provided short but high-count flat-field exposures. The filtered LFC was delivered to the pre-slit area of UHRF with a single-mode optical fibre and launched through a rotating diffuser onto the spectrograph slit – see Figure 2.

The LFC test with UHRF was very successful. Part of an example exposure is shown in Figure 3. UHRF was set to 780 nm in the full resolution mode ($R = 940$k), allowing ~1 nm of a single echelle order to be recorded with the MITLL3 2kx4k CCD. The CCD dewar was rotated so that the spectral direction coincided approximately with the CCD rows. Figure 3 shows the expected dense series of LFC modes in the spectral direction; more than 500 are recorded in just ~1 nm! By plotting a single row, one can see the very high signal-to-noise ratio obtained in this 20-s exposure. The comb modes themselves are unresolved, so each recorded mode is a replica of the instrumental profile (IP), and this will facilitate the IPSV measurement. Indeed, one can even see by eye in Figure 3 that the IP is non-Gaussian.

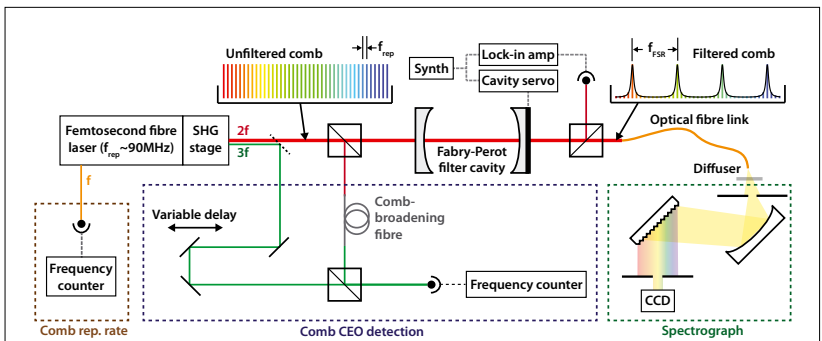

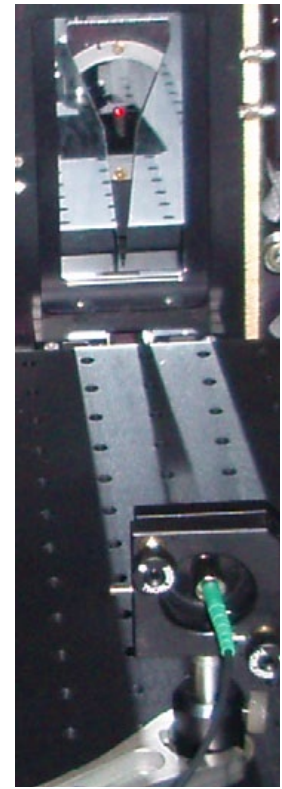

**Figure 2:** Launching the LFC into UHRF in the pre-slit area. The mode-filtered LFC light at 780 nm is delivered via single-mode optical fibre (black, insulated cable with green, mounted end plug) to the pre-slit entrance of UHRF and launched in free space through a diffuser (not shown) onto the slit (reflective area with decker jaws), greatly over(under)-filling it in the spectral(spatial) direction. The red laser spot visible in this photograph is not the LFC, which was at 780 nm, but a diode laser at 638 nm fed by the optical fibre from the LFC set-up bench for alignment purposes.

**Figure 1:** Laser frequency comb set-up with AAT/UHRF. The comb is fully specified by the 'tooth' spacing (repetition rate) and carrier envelope offset (CEO) frequency. The former is measured directly with a photo-diode and counter. The second harmonic generation (SHG) stage frequency-doubles the femtosecond-pulsed infrared laser (1.56 μm) to 780 nm (2f) and also produces some third harmonic light (3f) for the CEO measurement: the high-intensity, 780 nm light has its fairly narrow bandwidth broadened in a non-linear micro-structured crystal fibre for mixing with the 520 nm (3f) light; the 'beat note' specifies the CEO frequency. Ten out of every 11 comb modes are interferometrically suppressed using a Fabry-Perot cavity, increasing the effective mode separation from 94 MHz to 1 GHz, wide enough to be resolved from each other by UHRF in its R = 940k mode. The cavity's free spectral range (FSR) is locked to the filtered comb's effective repetition rate by the servo electronics.

Preliminary 2-D modeling of individual comb exposures has already revealed that IPSVs in the spectral direction are small, less than ~5%. Since the MTLL3 is a deep depletion device, this is not surprising. However, to our knowledge, this has not been measured (or, at least, reported) before. We also recorded a series of exposures where the repetition rate was varied slightly between exposures, effectively stepping the comb along the CCD, which may yield alternative constraints on IPSVs. Finally, the stability of UHRF during our test will be established from a series of short exposures, something not possible with the usual thorium calibration method because, with UHRF's very high resolution, many minutes are required to obtain adequate photon counts in just a handful of lines. Full results will be reported in detail elsewhere.

This LFC demonstration would not have been possible without the expertise and all-round resourcefulness of many AAO staff, particularly Steve Lee, Rob Dean, Stephen Marsden, Doug Gray, Steve Chapman, John Collins and Winston Campbell. ✦

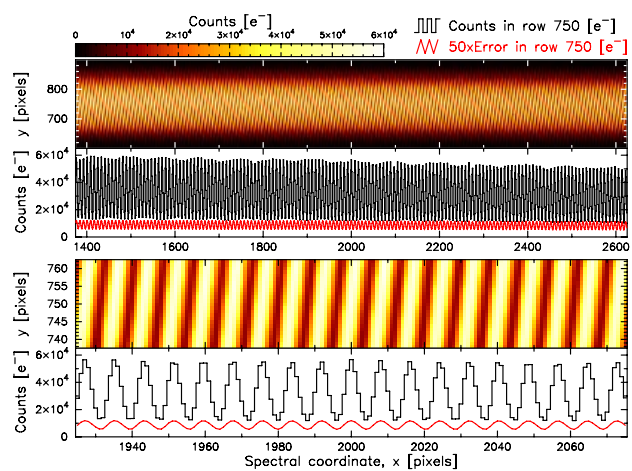

**Figure 3:** Example exposure (bias-corrected and flat-fielded) at 780nm from the LFC test with UHRF. The lower panels show a zoomed-in portion of the upper panels. Below the CCD image is a spectrum (black histogram) extracted from a single row of the image with the error spectrum (red line) exaggerated by a factor of 50. The comb modes provide dense information for both calibration and detailed characterization of various systematic errors in both the spectrograph and CCD. For example, inspection (try squinting) of the lowest panel shows an asymmetric instrumental profile (IP). Detailed modeling suggests a double-Gaussian IP, consistent with that found using the UHRF HeNe alignment laser (Diego et al. 1995). Note that the spectral direction is tilted with respect to the CCD columns. This is very important for measuring intra-pixel sensitivity variations (IPSVs): pixels in the same column but different rows have different flux distributions across them in the spectral direction so, assuming they have the same (or highly correlated) IPSV maps, the apparent centroid of a comb mode will shift anomalously along a column.

# Galaxy And Mass Assembly (GAMA): Successes, Progress and Plans

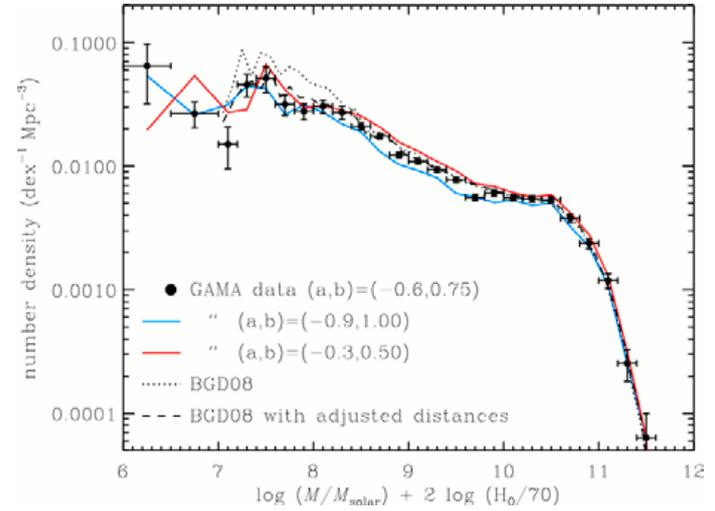

An initial estimate of the galaxy stellar-mass function (Baldry et al, in prep), demonstrating the extremely low masses to which GAMA is sensitive, $10^6$ M$_\odot$, along with a confirmation of the steep rise in numbers at the low mass end (Baldry, Glazebrook, Driver, 2008).

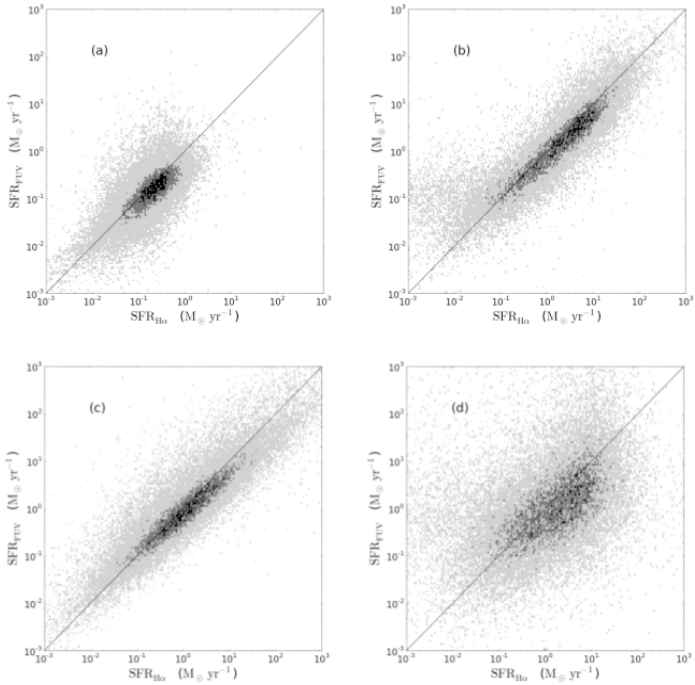

Comparison between far-UV and Hα-derived star formation rates, for different approaches to obscuration correction (Wijesinghe et al., 2011b). This illustrates the effect of using either Balmer decrement or the UV spectral slope, β, in the corrections. (a) No correction; (b) Both corrected using Balmer decrement; (c) Both corrected using β; (d) FUV corrected using β, and Hα corrected using Balmer decrement. It is evident that systematics in the measurement of β prevent it from being a robust obscuration metric.

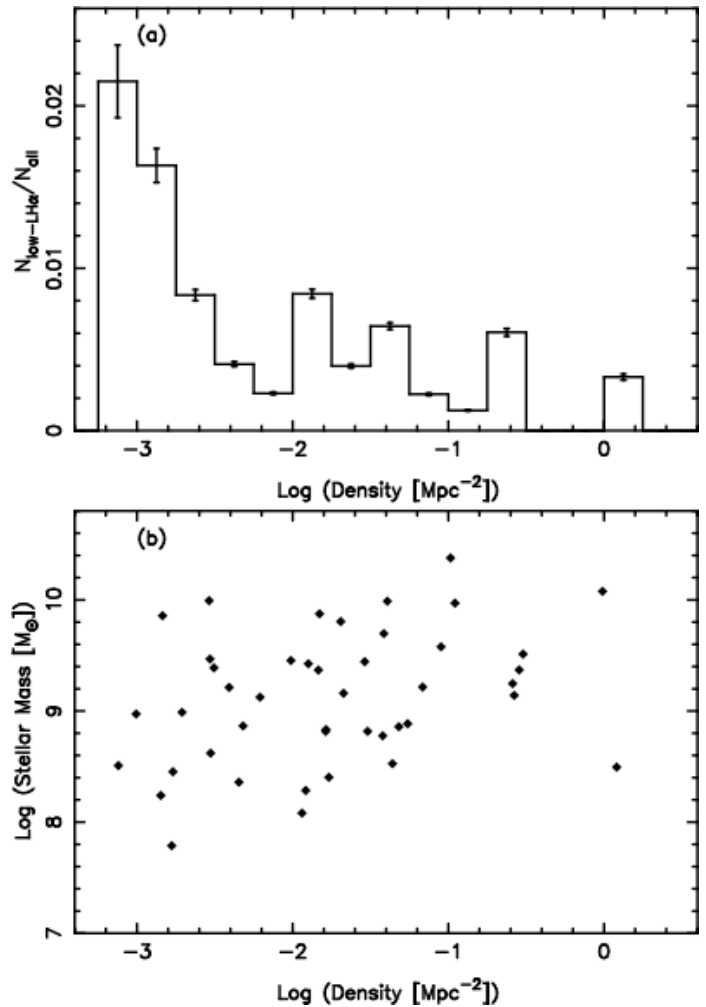

The local galaxy density for a population of low Hα luminosity (L$_{H\alpha}$ < $4 \times 10^{32}$ W) star forming galaxies in GAMA (Brough et al., 2011). This highlights that these low mass, low star formation rate systems preferentially populate low density, void-like regions, and comprise an increasing proportion of the population at progressively lower densities. They are completely absent from the highest density regions (not shown, an order of magnitude more dense than the highest level on this figure).

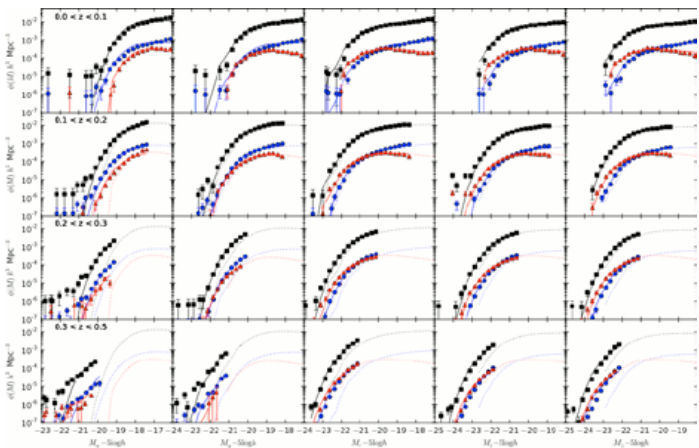

The multiwavelength optical (ugriz) luminosity functions for the full GAMA sample (Loveday et al., in prep). The sample has further been split by both redshift and optical colour, with redshift increasing from top to bottom, and both blue and red galaxy populations shown. The solid lines are parametric evolving luminosity functions, while the symbols show the SWML estimates. The dashed lines reproduce the lowest redshift LFs to highlight the evolution in the higher redshift bins.

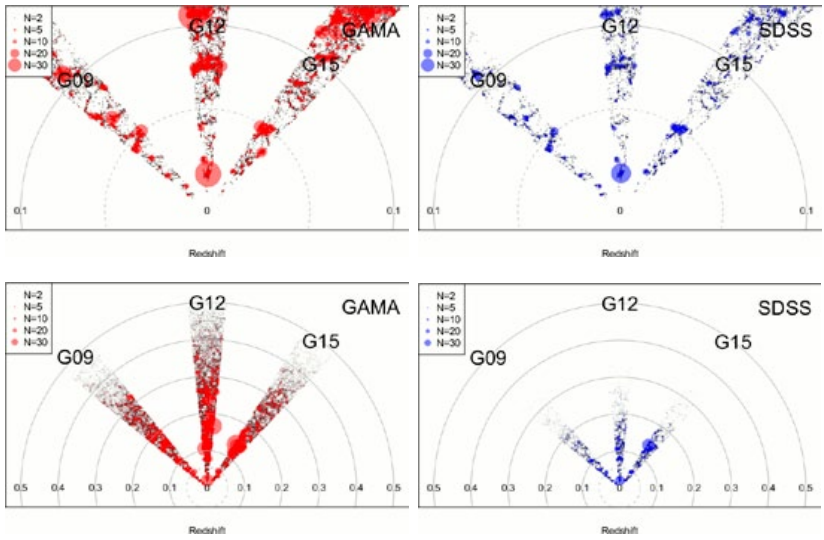

Cone plots illustrating the distribution within GAMA (left) and SDSS (right) of galaxy groups (Robotham et al., in prep). The multiplicity of the groups is represented by the size of the coloured circles. The top row shows groups out to z=0.1, the bottom row shows groups out to z=0.5. The increased depth of the GAMA survey is evident in the larger number of groups detectable.

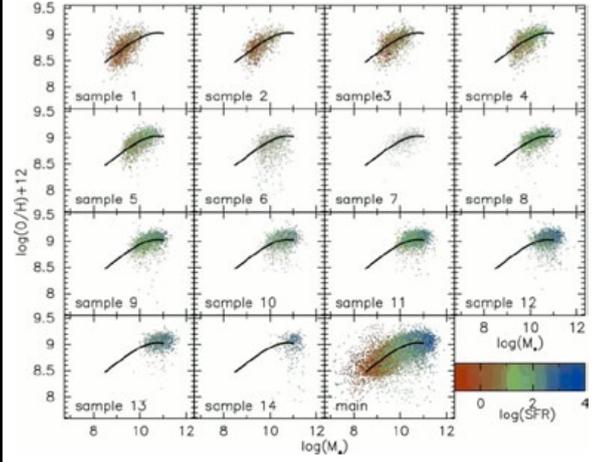

The mass-metallicity relation for GAMA galaxies, colour coded by star formation rate, for a sequence of volume-limited samples in narrow redshift bins (Foster et al., in prep). The solid line on each panel is the same in each case, and shows the local mass-metallicity relationship found by Kewley and Ellison (2008) from SDSS data. There is no evidence of evolution with redshift (out to the z=0.35 limit sampled here) in the mass-metallicity relationship, nor of strong variations with star formation rate.

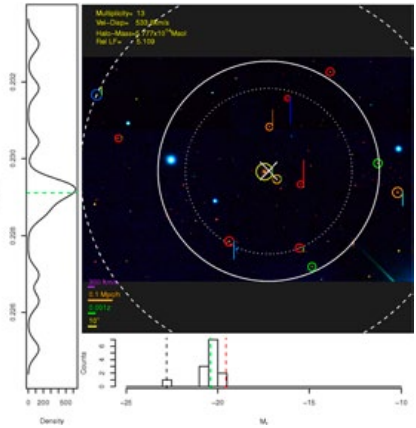

An example galaxy group identified from the GAMA survey (Robotham et al., in prep). The luminosity weighted centroid of the group is marked by the cross. The bars show the redshift of group members with respect to the group velocity, and the colours of the circles around the group members represent galaxy u-r colour.

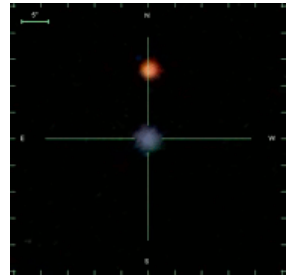

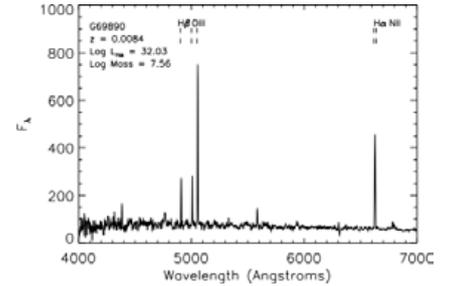

An example of the lowest Hα luminosity objects found in GAMA, illustrative of the blue dwarf morphology of most of this sample (Brough et al., 2011). This galaxy, at z=0.008, has a SFR of about 0.01 $M_\odot yr^{-1}$, and a mass of 3x$10^7$ $M_\odot$.

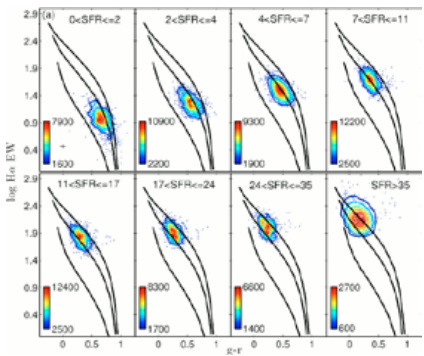

A diagnostic diagram showing evidence for variation in the stellar initial mass function (IMF) between galaxies of different star formation rate (Gunawardhana, et al., 2011). The Hα equivalent width as a function of g-r colour is shown for this volume-limited sample from GAMA. The solid lines show model predictions for different high-mass slopes of the IMF, with a Salpeter ($\alpha$=-2.35) slope in the middle, a flatter slope ($\alpha$=-2) above, and a steeper slope ($\alpha$=-3) below. The data are split by star formation rate, and the colour shows the density of the data points in the diagram. It is clear that higher SFRs appear to be better associated with flatter IMF slopes.

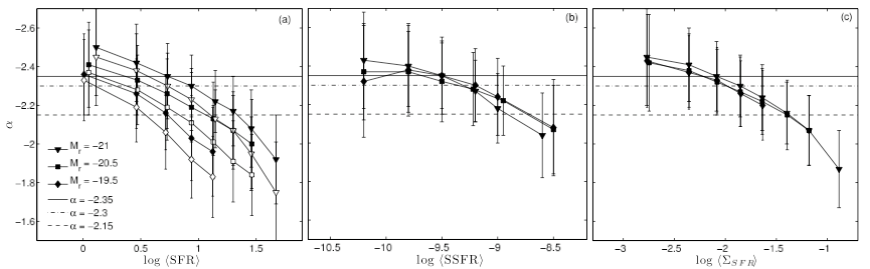

The inferred IMF slope, $\alpha$, for three independent volume limited samples within GAMA (Gunawardhana et al., 2011). (a) As a function of SFR, with two different obscuration curves used in making obscuration corrections; (b) As a function of specific SFR; (c) As a function of SFR surface density. The underlying IMF slope dependency seems to be related primarily to the local density of star formation, as parameterised by surface density of SFR, or specific SFR.

# The Search for Planets Around Detached Eclipsing Systems (SPADES)

Simon O'Toole (AAO), Tom Richards (Variable Stars South), Scott Thomas (University of Canterbury) and the SPADES team.

http://bit.ly/SPADES_AAO

The diversity of exoplanet properties uncovered in the last 15 years presents tremendous challenges to our understanding of planet formation and evolution. There have been many new surprises, from planets orbiting pulsars, to the under-dense "hot Jupiters", to planets with highly eccentric orbits. One of the most surprising discoveries has been the significant number of planets and substellar companions found in binary star systems: approximately 17% of exoplanetary systems discovered to date contain at least two stars (Mugrauer & Neuhäuser 2009). Most of these systems contain planets orbiting a Main Sequence star with a companion at large distance – 250 to 6000 AU – however three systems have been discovered where the binary separation is small: less than 40 AU (e.g. Hatzes et al. 2003). These are sometimes referred to as "S-type" ("satellite-type") planetary systems. Perhaps the most remarkable discoveries though, have been those of systems where the planets orbit *both* stars in the binary system (e.g. Lee et al. 2009); these are the "P-type" ("planet-type") or circumbinary planets. The questions to ask now are: how common are these systems? In what environments do we expect to find circumbinary planets and how do they form?

## Formation and Stability

The detection of circumbinary planets orbiting close stellar binaries provides us with an important way to test and constrain planet formation models. While the detection of planets around single stars has led to a paradigm shift in our understanding of planet formation, there are still several different models that can produce effectively the same planetary system configurations (e.g. Ida & Lin 2004). By investigating how stellar multiplicity affects planet formation we can ask questions such as: does binary star formation promote or hinder planet formation?

What is expected is that in a binary system the planet-forming nebula will be disrupted by the dynamics of the two stars. The perturbations of the central binary star will lead to the substantially different evolution of a protoplanetary disk compared with that of a single star (Artymowicz & Lubow 1994). Pierens & Nelson (2008) investigated the formation, evolution and migration of planets in circumbinary discs. In binary systems where the initial separation of the two stars is 1 AU, they found that Saturn-mass circumbinary planets are likely to be quite common, while Jupiter-mass planets will be less common, and exist at larger distances from the binary. There appear to be *no* formation studies of systems where the binary orbital periods are on time-scales of days or tens of days. Holman & Wiegert (1999) studied the long-term stability of planets in binary systems over a range of mass ratios and binary eccentricities. They found that planets beyond 3.7 times the binary separation should be stable.

## Currently known systems

Up to now, all circumbinary planets or brown dwarfs have been discovered orbiting systems containing either a white dwarf or hot subdwarf. The secondary star is an M star in a short-period orbit in each case (e.g. HS0705+6700, Qian et al. 2009). There is a strong selection effect at work here, however: each of these systems have been discovered by teams studying (pre-)cataclysmic variables rather than searching for circumbinary planets (at least initially). All systems have been discovered using eclipse timing: the times of each primary eclipse minimum are compared in an observed eclipse times minus the expected or computed eclipse times (O-C) analysis, and periodic signals that cannot be explained in any other way, such as magnetic effects or angular momentum loss (Applegate 1992), must be due to third bodies.

It is unclear *when* these circumbinary planets formed: did they form as "first generation" planets, at the same time as the progenitor star; or did they form *after* a vast amount of mass was lost from the system through a symbiotic or common envelope phase? Perets (2010) has suggested that there may enough material in, e.g., an accretion disc around a symbiotic star, to form planets, and has used this idea to explain the known systems. There is some observational evidence that protoplanetary discs can form around these systems (e.g. Ireland et al. 2007). Detection of – or stringent limits on – a planetary system around a close Main Sequence binary would be an important step towards answering these questions.

## Circumbinary discs

Can circumbinary planets form around binary systems where *both* stars are on or near the Main Sequence? Circumbinary discs have been detected around young spectroscopic binaries such as CoKu Tauri/4 (Ireland & Kraus 2008), HH 30 (Guilloteau et al. 2008) and GG Tau. In the latter system, the disc has been resolved and an inner disc cavity has been observed, which is due to the tidal torques exerted by the central binary (Dutrey et al. 1994). Brightness asymmetries have also been seen in subsequent HST imaging, which suggests the disc is warped by tidal effects (e.g. Krist et al. 2005).

It therefore appears that the necessary factors (discs and stability) are there for circumbinary planets to form and survive around binary systems containing Main Sequence stars. Detecting such planets is now the final piece in the puzzle.

## Previous studies

There have been a limited number of attempts to detect such systems, mainly because standard planet search methods are difficult to implement. The Doppler velocity method used by planet search teams such as the Anglo-Australian Planet Search (e.g. O'Toole et al. 2009) cannot easily be applied to double-lined spectroscopic binaries. Konacki et al. (2009) initiated a radial velocity search for circumbinary planets around Main Sequence binaries using a novel combination of the standard method and the two-dimensional cross-correlation technique TODCOR introduced by Zucker & Mazeh (1994). To date, they have managed to achieve precisions down to $\sim$2ms$^{-1}$ for the individual components, but significantly higher (> 10 times) for the systems overall. They detected no candidate exoplanets based on the examination of 10 double-lined systems with periods longer than ~5 days. It is uncertain however, whether this technique will be precise enough to detect the tiny variations in the systemic velocities caused by circumbinary planets.

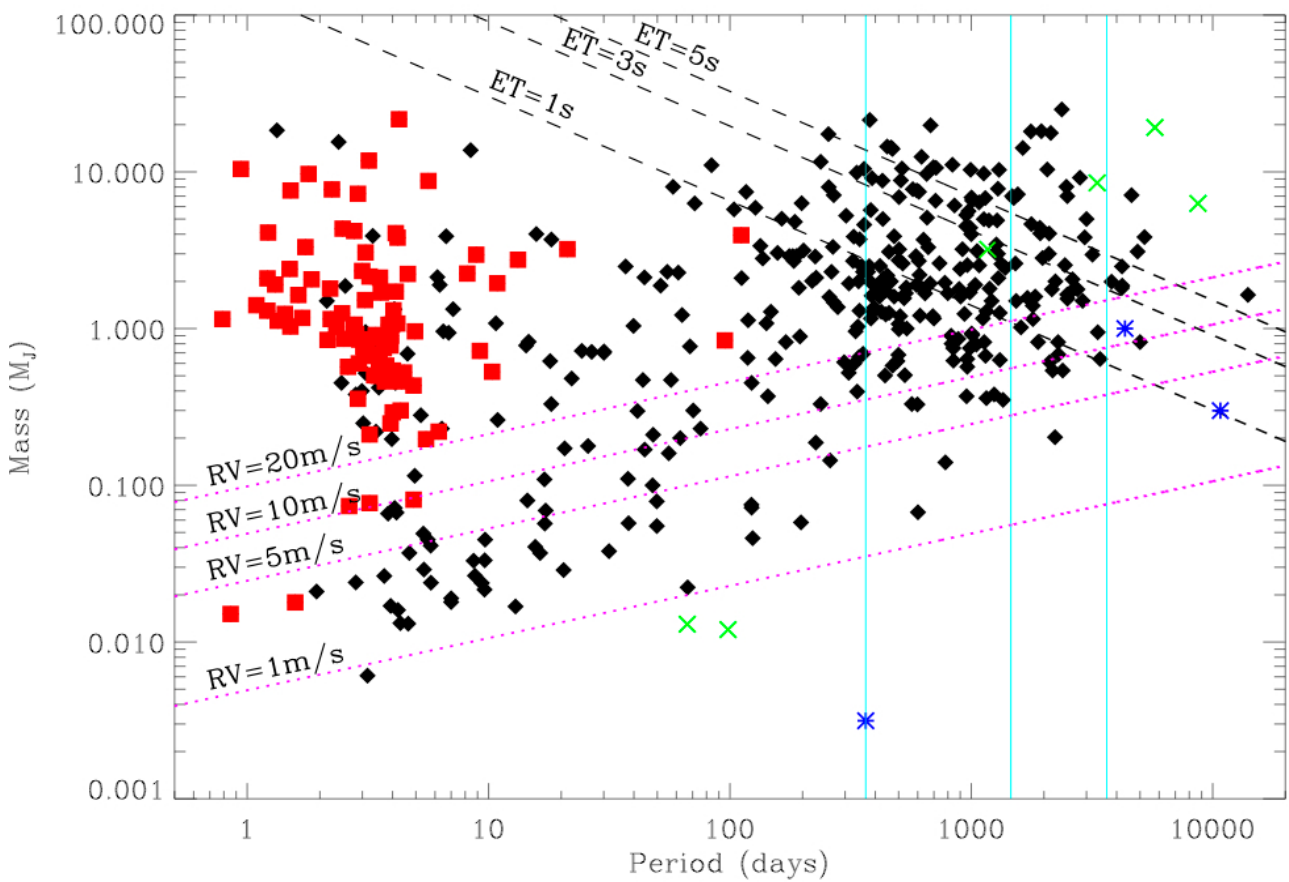

**Figure 1:** Discovery space for circumbinary planets around a binary system with a total mass of 1.5M$_\odot$. The dashed lines marked ET represent various timing accuracies, while the magenta dotted lines marked RV represent various Doppler velocity accuracies for a star with one solar mass. Also shown are all known Doppler-detected planets (black diamonds), transiting planets (red squares), and timing-detected planets (green crosses), along with Earth, Jupiter and Saturn (blue asterisks). The cyan lines are marked at 1, 4 and 10 years.

The eclipse timing method does not suffer from these problems. If the target systems are properly selected, it therefore provides the best chance to detect planets around the large number of eclipsing binaries with periods less than 5 days. *For this reason, we have started the SPADES project.*

### The SPADES project

The Search for Planets Around Detached Eclipsing Systems (SPADES) project is designed to look for substellar companions orbiting around eclipsing binaries containing Main Sequence or Sub-giant stars with spectral types from A to M. It is a collaboration between amateur and professional astronomers. We will use the eclipse timing method discussed above and involve multiple telescopes at different sites. The eclipsing systems have been selected from the General Catalogue of Variable Stars (GCVS – Kazarovets et al. 2009), have orbital periods between 1 and 5 days, and are Algol-type binaries where the stars are detached; i.e. there is no mass transfer between them. By selecting targets in this way, we are ensuring that the eclipse minima are sharp and can be measured with high precision, and the eclipse durations are short enough that a whole eclipse can easily be observed in one night. Other periodic and quasi-periodic variations – due to magnetic variations and angular momentum loss, for example – will also be minimised. To achieve adequate phase coverage for our *O–C* analyses, we are collaborating with several experienced amateur astronomers through the Variable Stars South network (see http://www.variablestarssouth.org), and plan to work with undergraduate student programs where appropriate. Variable Stars South has members across the world. (Note that the conditions of joining the SPADES network are access to a telescope with a minimum aperture of about 30cm, a CCD camera, and an accurate method of calibrating and recording observing times.)

The project will last for at least five years, as, unlike Doppler velocity planet searches, eclipse-timing measurements become *more* sensitive to lower masses at longer periods. This can be seen in Figure 1, where we show the discovery space for circumbinary planets using the eclipse timing method. The timing precision thresholds assume a binary system mass of 1.5M$_\odot$, equivalent to a late G-type plus K-type Main Sequence binary; they are calculated using equation 5 of Sybilski et al. (2010). The green crosses in Figure 1 show the currently known planets detected using the timing method. The two crosses at the bottom of the figure near 100 days are the pulsar planets around PSR 1257+12 (Wolszczan & Frail 1992), while the other systems are orbiting (pre-)white dwarfs, as discussed above. Also shown in the figure are sensitivity levels for four different Doppler velocity precisions for comparison. Finally, the data points for currently known exoplanets are taken from the Exoplanets Encyclopaedia (see http://exoplanet.eu).

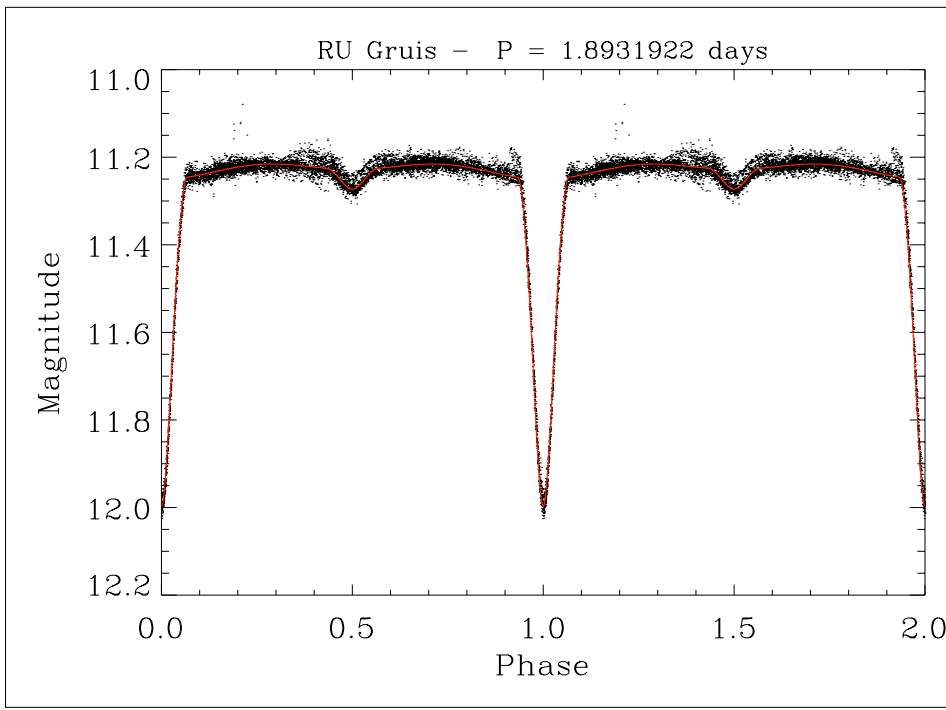

**Figure 2:** SuperWASP light curve of the eclipsing binary RU Gruis, phased with a period of 1.8931922 days and the time of minimum light. Overplotted in red is the best-fit model light curve calculated using John Southworth's JKTEBOP code. We are using this code to measure the light elements of each system. The updated ephemerides will be useful for both the SPADES project and the eclipsing binary community.

## Current status and future directions

Of course, observing eclipsing binary stars is not a new thing. There are several online archives such as the *O–C Gateway* (see http://var.astro.cz/ocgate/index.php?lang=en) that contain more than 50 years of observations for some of the SPADES targets; however, much – but not all – is low quality and based on visual or photographic observations. In fact, the majority of eclipsing binary systems in the southern sky are poorly studied, with little known about their spectral types, component masses and other properties. One part of SPADES that is of great interest to the amateur astronomy community (in particular), is the characterisation of these systems using spectroscopy; this is one of the side goals of the project.

An important starting point for the SPADES project is to search for already existing observations of our targets. One of our AAO Summer Vacation Scholars, Scott Thomas, is examining various online data archives such as WASP (Wide Angle Search for Planets, see http://www.wasp.le.ac.uk/public/) and ASAS (All Sky Automated Survey, see http://www.astrouw.edu.pl/asas/?page=main), to see which of our targets have been observed enough times to measure accurate ephemerides. He is fitting the light curves and deriving parameters using JKTEBOP (Southworth et al. 2007), and we plan to publish a catalogue of these values, focusing on interesting objects, of which there are several.

An example of Scott's work in Figure 2 shows the partially eclipsing system RU Gru; in this case, the minimum eclipse time is accurate to ~4 seconds over the approximately two years of data. The aim of SPADES is to achieve an eclipse timing precision of approximately 1 second, which would allow us to detect a Jupiter-like planet in a Jupiter-like orbit, however, the eclipses of some of our targets may not be sharp enough to achieve this, so we have also plotted lower precisions in Figure 1. Even with timing precisions of 5 seconds, we will still detect companions with only a few Jupiter masses in four years. Jupiter-mass planets are more likely to be formed and stable at these kinds of periods (Pierens & Nelson 2008). With higher S/N ratios from larger telescopes (the WASP cameras are only 11.1cm in diameter), we are confident of achieving our target precision of 1 second for the majority of systems. It is possible that the accuracy of the eclipse times will also be improved by examining individual eclipses. Looking for *O–C* variations in WASP data is also part of Scott's project.

The SPADES project will help to answer fundamental questions about planet formation, circumbinary discs and stellar multiplicity. It will also provide an important resource for both observational and theoretical stellar astrophysics in the form of a large database of fundamental parameters of close binary stars. Finally, it will create strong links between the amateur and professional astronomy communities in Australia.

# The NASA EPOXI mission flyby of comet 103P (Hartley 2)

Malcolm Hartley (UK Schmidt Telescope, AAO)

I had seen many images of this comet over the weeks preceding the mission fly-by on the morning of November 4 2010 but it is impossible to describe the palpable anticipation at Mission Control of the Jet Propulsion Laboratory early that morning. There was a big cheer and much back slapping (but no image as yet) minutes after the fly-by indicating that communication with the spacecraft was still OK. There had always been the possibility that debris surrounding the comet could have impacted with the spacecraft and with a relative fly-by speed of 12 kilometres/second it wouldn't take much to terminate the mission. The first close encounter images took a further 20 minutes or so to download. Final flight corrections in the days preceding the encounter had been carefully executed using the latest updates to the position of the comet from radar observations made with the Arecibo facility in Central America. These observations indicated that the comet was markedly elongated and possibly "peanut" shaped. Don Yeomans, a veteran of comet and asteroid orbit/ephemeris predictions was interviewed on NASA TV shortly before the first image appeared. He was handling shelled peanuts at the time. The first image raised an even greater cheer and gasps of excitement from one and all as the comet was clearly peanut shaped, outgassing actively and definitely something the Science Team as well as the Rocket Scientists were overjoyed to see. I was pretty happy too!

Comet 103P was discovered at the UK Schmidt on a plate taken for the Equatorial Red survey in March 1986. During quality control of the plate, I was the lucky person who discovered the image, which has since been described by David Malin as a "mere smudge" on a photographic plate. It was hardly even that! The discovery image had a faint halo around the trailed image which is the telltale sign that it isn't an asteroid. Follow up images obtained a few days later confirmed that the object was a comet and the orbital elements indicated that it was "new" with a period of just over six years so as well as the catalogue number of 103P it was also named Hartley 2.

In early 2009, a reporter with New Scientist magazine contacted me to ask how I felt about the EPOXI mission targeting Hartley 2 for a fly-by in 2010. (Hartley 2 was in fact the back up target, the original target, comet Boethin was lost). Since then, that has been the most commonly asked question in relation to the comet! When I heard the news it did occur to me that in October/November 2010 the event might make the local press, especially with the prospect of the comet becoming a naked eye object. I could never have imagined the extent of personal involvement in the event nor the sense of attachment which would be constructed by the process.

In mid 2010 I was formally alerted of the fly-by with an email from Ron Baalke, an astronomer based at JPL asking if I was the Malcolm Hartley who had made the discovery. One thing led to another and the outcome was an invitation from the outreach/media team at JPL to be present at Mission Control for the fly-by and also be available for public talks, lectures and media interviews. Laura, my wife, agreed to come too and so we planned for a holiday/business trip of a couple of weeks.

We were under the JPL umbrella for the last 9 days. They were extremely welcoming and generous, to the point of embarrassment at times. We were taken out to dinner with a group of project managers from the Laboratory along with the Science PI, Mike A'Hearn from the University of Maryland and the menu at the venue was headlined, "Welcome Malcolm Hartley". Street banners in the JPL complex, logo emblazoned gifts and all staff briefings put the fly-by event at the front of JPL life.

The first public engagement was a talk to be given at the Griffith Observatory in Los Angeles. As the deputy project manager for the mission Don Sweetnam drove us there he remarked that it was a small public observatory and the event would be low key. It wasn't small! With a staff of over 20, a budget of millions, a planetarium and working telescopes to boot it was definitely not small and the auditorium, the Leonard Nimoy Event Horizon Theatre, was huge and to top it all I was billed as the "Comet Hunter" speaker. I'm not used to cameras, microphones, memory sticks and giant screens and worst of all I wasn't a comet hunter, and never had been. The media tutorials I'd had in Sydney before coming now came into their own; stick to the script and squeeze in as much about comets as you possibly can. It seemed to work. Pictures of the Warrumbungles and the world's biggest virtual solar system drive as well as some hairy site-testing stories from the 1970's were enough to distract as well as intrigue the audience about the diverse experiences an astronomical career can deliver.

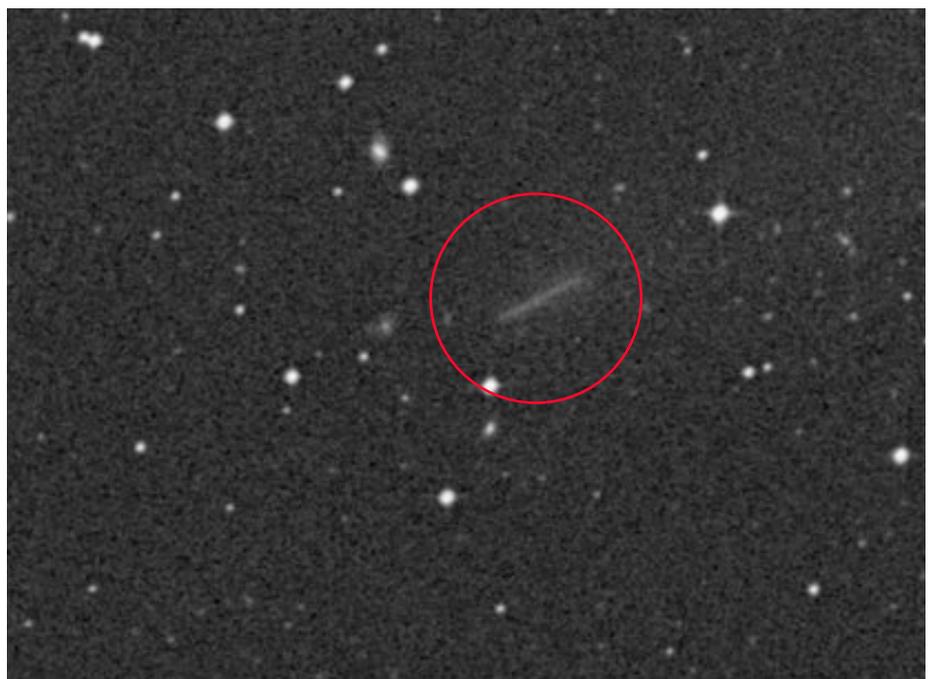

**Figure 1:** Discovery image of Hartley 2

A further two talks at JPL to the Rocket Scientists and the Pasadena Public completed the programmed element of the trip, with the addition of an address to a few hundred school children who were at JPL on the day of the fly-by. Equipped with a NASA cell phone as well as my aussie mobile, media queries and interviews could and did occur at all times of day and night.

I had imagined JPL was much larger than the AAO, perhaps with a staff in excess of five hundred. There are over seven and a half thousand people working there! We had a minder wherever we went in the complex as security is paramount. We were introduced to the Director and two deputy directors, we saw the next Mars Rover under construction and had the privilege of attending the "lock out" science meetings which were held in the days leading up to the fly-by to thrash out all things associated with the comet and hoped-for outcomes, a truly extraordinary experience.

The enthusiasm of the Science Team as well as the Rocket Scientists and general JPL population was infectious. We loved the way they celebrated and promoted the science. They kept calling it my comet and by the time of the fly-by with the accumulation of congratulations, endless handshakes and introductions I almost believed it was. We really felt like honoured guests and the whole experience was as fantastic as it was unexpected. Thank you JPL and NASA and in particular, Aimee, Judy, Beatriz, Don, Tim, Ron, and David (and Helen Sim at the AAO).  ✦AAO

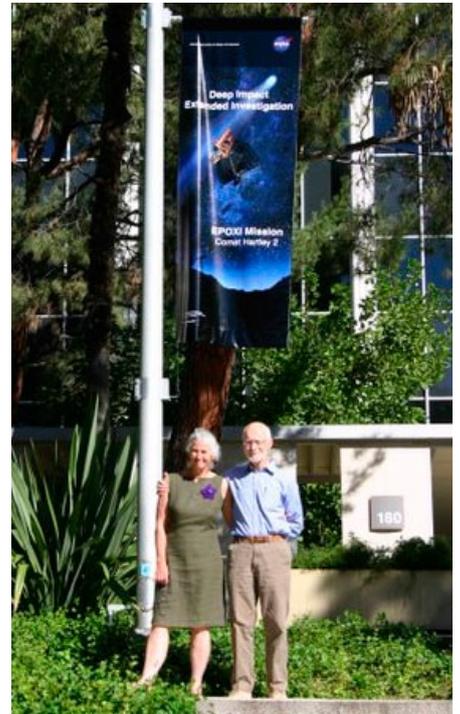

**Figure 3:** Malcolm and Laura Hartley below an EPOXI banner at JPL

**Figure 2:** The first fly-by image of Hartley 2

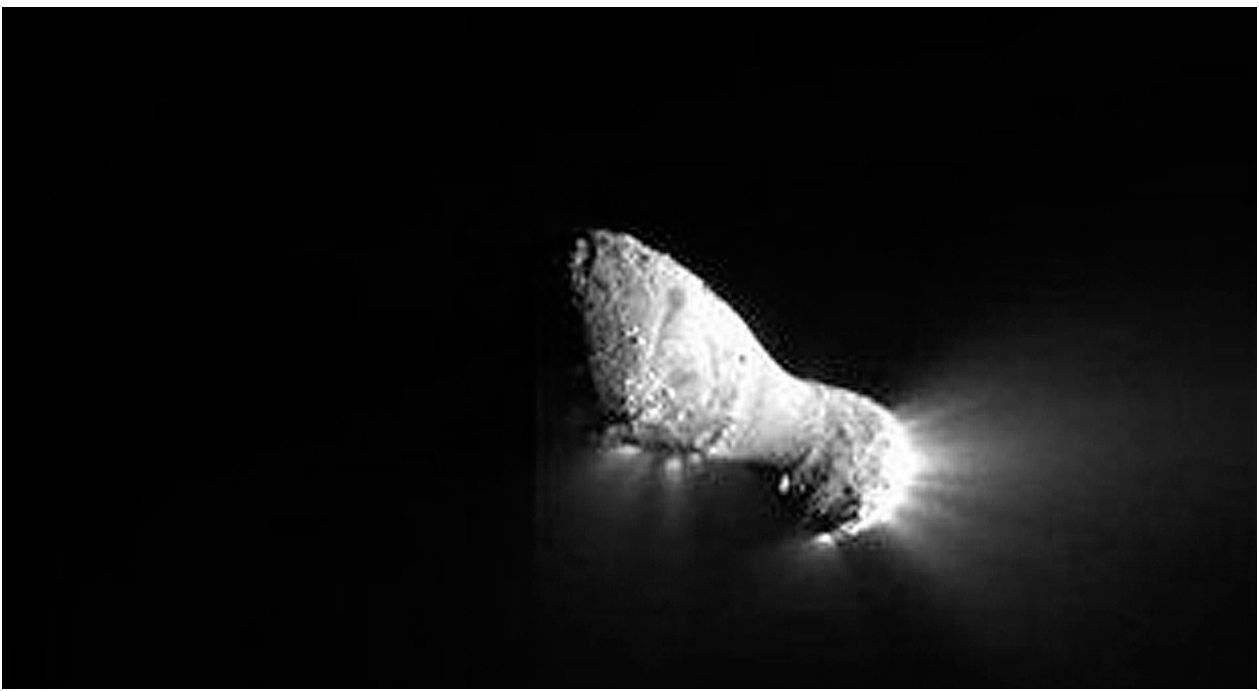

# Southern Cross Astrophysics Conference Series*
# **2011 Conference Announcement**

Chris Lidman (AAO) on behalf of LOC and SOC

As part of the the Southern Cross Astrophysics Conference Series*, we are pleased to announce a conference, **Supernovae and their Host Galaxies** to be held in Sydney, Australia, 20-24 June 2011

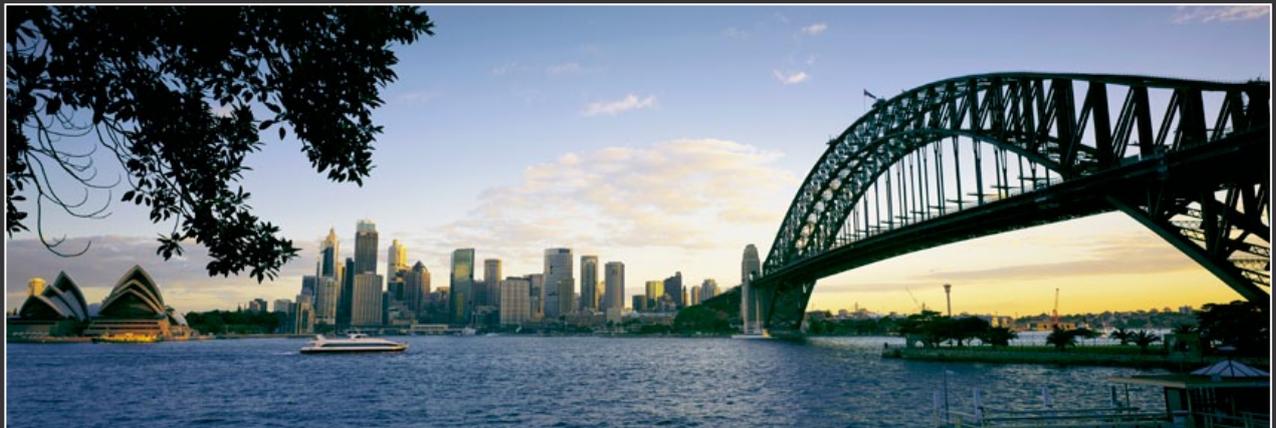

Photo Credit: Sally Mayman and Tourism NSW

## Motivation

The current generation of wide field transient surveys will revolutionise our understanding of why stars become supernovae. Designed to revisit large areas of sky at multiple wavelengths, these surveys are now discovering hundreds of supernovae each year. During the coming year, the number of supernova discoveries will increase even further as new transient surveys come online. As well as finding rare and possibly new types of supernova, these surveys will generate new insights into both core-collapse and thermonuclear supernovae. It is therefore timely to have a conference that explores the current (observational and theoretical) supernova landscape and the connection between supernovae and their host galaxies.

Some topics to be covered include:

- the many different paths to a supernova explosion
- the progenitors of supernovae
- supernova remnants
- supernova rates
- the relationships between the properties of supernovae and the properties of their host galaxies
- supernovae as tracers of star formation
- unusual supernovae
- unexplained transients
- current and future transient surveys

**To find out more and to register, visit http://www.aao.gov.au/AAO/southerncross/ Registrations open February 1st.**

* Fourth annual conference, jointly Organised by CASS and the AAO

# AUSGO CORNER

Stuart Ryder (Australian Gemini Office, AAO)

## Proposal Statistics

For Semester 2011A ATAC received a total of 29 Gemini proposals, of which 13 were for time on Gemini North, 2 were for exchange time on Keck or Subaru, and 14 were for time on Gemini South. The overall oversubscription (1.58) was the lowest since 2006A, which can be attributed in large part to ongoing delays in commissioning of FLAMINGOS-2, GSAOI with MCAO, and the new red-sensitive CCDs on GMOS-North. The return of GNIRS on Gemini North did help trigger a few new proposals. At the ITAC meeting Australia was able to schedule 19 programs into Bands 1–3, 8 of which involved joint allocations with other Gemini partners. For Magellan we received 9 proposals, keeping Magellan oversubscription very healthy at 3.13, with MIKE and the various modes of IMACS continuing to be the most popular requests.

In 2010A, all but one of the Band 1 programs (and one with rollover status) were completed; 5 of 6 Band 2 programs got 85% or more of their data; and half the Band 3 programs were also completed. Barely half of Magellan time was usable in 2010A due to unusually poor winter weather.

## GNIRS System Verification

With the return to service of the new-and-improved Gemini Near-InfraRed Spectrograph (GNIRS) on Gemini North late in 2010, a Call for System Verification (SV) observations was issued by the Gemini Observatory. This enables all the various modes of GNIRS (now including natural and laser guide star adaptive optics) to be tested on targets suggested by the community, and yield early science results from data with only a 2 month proprietary period. The 120 hours of available time was oversubscribed by a factor of 2.5. Three programs submitted by Australian PIs (Kerzendorf, Webster, and Landt) were selected and have been placed in the SV queue for potential execution in late-2010B/early-2011A.

## AGUSS

Each year since 2006 AusGO has offered talented undergraduate students enrolled at an Australian university the opportunity to spend 10 weeks over summer working at the Gemini South observatory in La Serena, Chile on a research project with Gemini staff. They also assist with queue observations at Gemini South itself, and visit the Las Campanas Observatory where the Australian Magellan Fellows are based. The Australian Gemini Undergraduate Summer Studentship (AGUSS) program is generously sponsored by Astronomy Australia Limited (AAL).

The selection panel was once again faced with a difficult task in selecting just two lucky recipients from 8 applicants. The 2010/2011 AGUSS recipients are Steven Saffi from the University of Adelaide, and Belinda Nicholson from the University of Melbourne (Figure 1). Belinda is working with Bernadette Rodgers analysing NICI data, while Steven is working with Percy Gomez on The Formation History of Massive Luminous Brightest Cluster Galaxies. They are due to present the results of their research via video link from Chile at the AAO/ATNF summer student symposium in Australia in mid-February 2011.

## Gemini Associate Director Visit

AusGO instigated and organised a visit to Australia by Andy Adamson (Associate Director for Science Operations) and Eric Tollestrup (Associate Director for Development) over 20–23 Sep 2010. They gave presentations to the community in Sydney (AAO), Canberra (RSAA) and Melbourne (Swinburne) on the transition plan for Gemini operations and on current and future instrumentation. They also held discussions with staff at AAO, RSAA, DIISR, OTAC members, and AAL.

## Fetch your OT Library!

Prior to the start of each new semester, Gemini releases an updated version of its Java-based Observing Tool (OT), to add new functionality and fix any bugs. Updating the OT is now as easy as clicking the "Update..." button when the OT is launched – this will check for any updates, download them, and install them the next time the OT is launched. Less obvious perhaps is that the OT Libraries for each instrument, which now contain extensive sets of template and example observations for practically any observing scenario, are regularly updated as well but not distributed with the core OT software. As with any observing program fetched by the user, a local copy of these libraries is kept on the user's own computer, and it is this which is displayed by default whenever the "Libraries" button on the OT menu bar is clicked. In order to ensure that you are using the most up-to-date Library for any instrument (even one you use regularly), we strongly recommend that users at least once a semester do a **File → Fetch Libraries...** and indicate which instruments they wish to refresh the locally-stored libraries for.

## Travel to Chile

As many regular users of Magellan and Gemini South will know, citizens of Australia, the USA, and Canada are required to pay a "reciprocity fee" on arrival at Santiago airport prior to passport control. This fee ranges from US$61 for Australians, to more than US$130 for US and Canadian citizens, and usually must be paid in US dollars cash. Once paid it is valid for the lifetime of your passport. Please check with the Chilean Embassy in Australia (http://chileabroad.gov.cl/australia/en) for the current fee and payment arrangements.

Note also that if you travel to Chile on LAN via Auckland, you may be required to obtain a transit visa for New Zealand, even if you are not planning to leave the departure gate area. Australian citizens and permanent residents do not require a transit visa, nor do citizens of countries for which a visa waiver applies (http://www.immigration.govt.nz/opsmanual/34163.htm)**.** Please check visa requirements before departure, or else you may not be allowed to board the flight in Sydney and you risk missing your observing run altogether!

## Magellan access and Fellows

AAL has signed a contract with the Carnegie Institution for Science for the purchase of a further 30 nights on the Magellan telescopes over two years, from Semester 2011B to 2013A inclusive. These nights are to be allocated by ATAC in the usual way. Regrettably it was not possible to negotiate a continuation of the Magellan Fellowship program, under which Australia paid for part of its access by seconding two Fellows to Chile for 2 years of observing support followed by a 3rd research-only year at an Australian institution of their choice.

The first of our Magellan Fellows, Dr David Floyd has recently completed his research posting at the University of Melbourne, and we are pleased to note that he has since had his position there extended. We also congratulate David on being selected as one of only 16 early-career researchers from across Australia for the 2010 "Fresh Science" event. This annual event brings together scientists with science communicators and enables them to learn effective communication skills, and give their research significant media exposure. David highlighted his recent results obtained using Magellan on the emission mechanism of lensed quasars. Our other inaugural Magellan Fellow, Dr Ricardo Covarrubias will shortly be finishing his research year at the AAO, and most likely taking up a postdoctoral position in the USA. We wish David and Ricardo every success, and thank them for their excellent service to AusGO and the Magellan user community.

## Science with the "Glowing Eye" contest winner

The judging criteria for the Australian Gemini School Astronomy Contest include both scientific merit and aesthetic value. While there is no doubt about the visual appeal of the winning entry for the 2009 contest, the "Glowing Eye" planetary nebula NGC 6751 proposed by Daniel Tran from PAL College (see the cover of the Feb 2010 issue of the AAO Newsletter), the same data has now proved to be scientifically valuable as well. A team led by David Clark at UNAM has carried out an extensive morphological and kinematical study of NGC 6751 (2010, ApJ, 722, 1260) combining data from the San Pedro Mártir 2.1m telescope, HST, Spitzer, and Gemini. The narrow-band Gemini images obtained for the contest are particularly valuable for delineating the various bipolar lobes and filaments, and separating the various stages of mass-loss from the progenitor AGB star from the ambient interstellar medium.

## Gemini Focus

Are you receiving a hardcopy edition of the twice-annual Gemini Focus magazine? If not, please let us know and we will add you to our mailing list, or you can access a PDF version at http://www.gemini.edu/index.php?q=node/27. The Dec 2010 issue features an article by AusGO's Dr Christopher Onken summarising just some of the amazing results notched up by the RSAA-built NIFS instrument in its first 5 years on Gemini North.

## New Magellan Proposal System

For the past two semesters, applicants for AAT time have been using a new web-based proposal preparation and submission tool developed by AAO software engineer Scott Smedley. This replaced the venerable but ageing web upload system that had been in use at the AAO since the late-1990s for the AAT, and since 2007 has also been used for Magellan proposals. With the commitment to maintain access to Magellan for at least four more semesters, AusGO has taken the opportunity to implement a similar web interface for Magellan proposals to that now in place for the AAT. From Semester 2011B onwards, Magellan proposals must be prepared using the interface at http://www.aao.gov.au/astro/apply/magsub/submission.php. Note that if you or any of your co-investigators have not previously applied for either Gemini or Magellan time through ATAC before, then you will need to e-mail us in advance to have them added to our database. ✦AAO

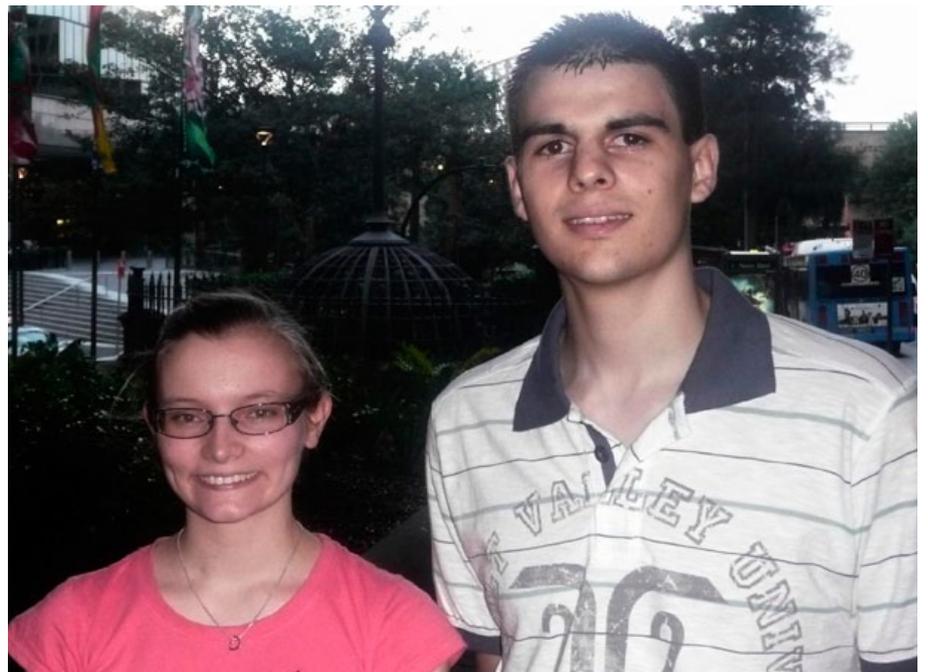

**Figure1:** AGUSS students Belinda Nicholson and Steven Saffi prior to their departure for Chile in Dec 2010.

# Australia's 2010 Gemini School Astronomy Contest

Christopher Onken (Australian Deputy Gemini Scientist,
Australia National University)

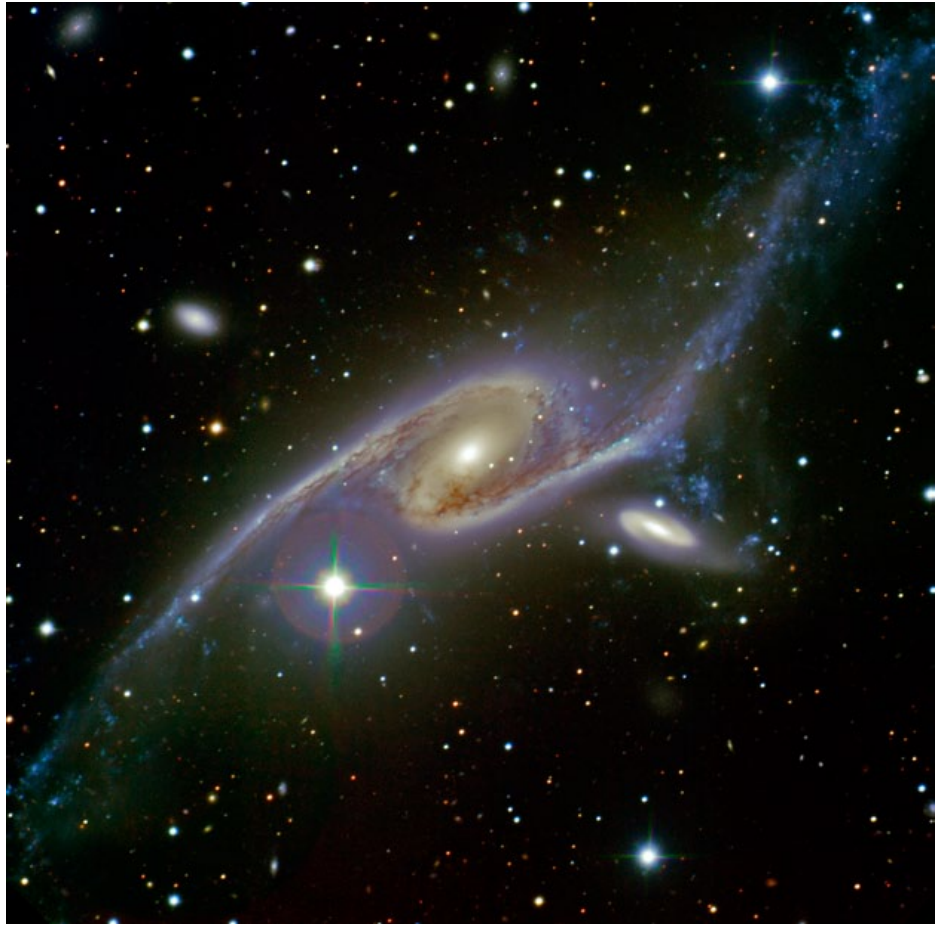

How would you spend an hour of Gemini time? That was the question students had to answer in the Gemini School Astronomy Contest. For the second consecutive year, high school students in Australia were given the opportunity to use the nation's largest optical telescope to take an image of something in the Southern celestial sky.

Budding astronomers from across Australia submitted their best ideas, having considered both the scientific interest of their chosen target and the potential for creating an inspiring picture. Balancing both the scientific and aesthetic merits of the entries, a group of volunteers from the fields of journalism, art, and astronomy selected the top three proposals.

The winning entry came from the Astronomy Club at Sydney Girls High School (SGHS), who suggested that Gemini examine the spectacular merging galaxy system, NGC 6872. Datasets in g', r', and i' filters were obtained with the GMOS instrument on Gemini South, and a colour composite was created by Travis Rector (University of Alaska Anchorage). The winning image appears on this Newsletter's cover.

The dramatic spiral arms of NGC 6872 extend roughly 60 kpc in each direction, making it several times larger than the disk of the Milky Way. The length of the arms and the smattering of blue light from newly formed stars are both indications that its smaller companion, IC 4970, is perturbing the galaxy. The SGHS image has captured this gravitational interaction at an early phase of the merging process.

In addition to the excitement of having a world-class observatory doing their bidding, the girls in the SGHS Astronomy Club won the chance to connect to Gemini by video. The students spent an hour linked up with Richard Valcourt, from Gemini's Public Information and Outreach (PIO) office, learning more about the observatory and its recent scientific results. The "Live From Gemini" event, hosted at ATNF's Marsfield headquarters, also featured introductions to Australia's other astronomical facilities by Stuart Ryder (Australian Gemini Scientist) and Rob Hollow (CSIRO's Education Officer). The AAO's Gayandhi de Silva and the ATNF's Bärbel Koribalski treated the girls to fascinating discussions of their respective research areas, as well as the paths they've taken to becoming professional scientists.

"Live From Gemini" sessions were also arranged for each of the two contest runners-up. I paid a visit to Benjamin Graham and his classmates at Whitefriars College (VIC), who were hooked up to Peter Michaud, Gemini's PIO Manager. Kieran Cerato and the Forest Lake College (QLD) Astronomy Club were hosted by the University of Queensland, and after their link to Gemini, University of Queensland's Holger Baumgardt and I gave the students a glimpse into the mysterious world of supermassive black holes.

As we begin 2011, the Australian Gemini Office is thrilled to be ramping up for another school contest. We thank ATAC for their generous allocation of an hour on Gemini South to continue building on our previous successes. This year, the eligibility has been broadened to include Years 5-12, giving us the opportunity to capture the imagination of even younger students and to help foster future generations of Australian astronomers. Entries are due by 13 May 2011.

http://ausgo.aao.gov.au/contest/

# AAO Distinguished Visitors

Andrew Hopkins (AAO)

# 2011
## AAO Distinguished Visitors

Beginning in 2010, the AAO now runs an annual Distinguished Visitor scheme. The aim of this scheme is to strengthen and enhance the AAO's visibility both locally and internationally, and to provide opportunities for AAO staff to benefit from longer term collaborative visits by distinguished international colleagues. The AAO is pleased to be hosting 5 Distinguished Visitors during 2011.

Each of our Distinguished Visitors will be giving colloquia both at the AAO and at other institutions during their visit, to highlight their work and collaborations with the AAO. They will also be engaging with the general public through public lectures or other outreach activities. We encourage members of the astronomical community to visit the AAO and take advantage of these visits by our distinguished colleagues to build new collaborations, to reinforce or rekindle existing ones, and to maximise the outputs from their time with us at the AAO. ◦AAO◦

**Details of the AAO Distinguished Visitor Scheme can be found on the AAO web pages at: http://www.aao.gov.au/distinguished_visitors/**

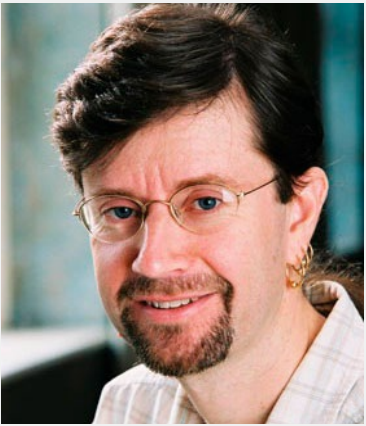

**Dr Peter Barnes**, (University of Florida), visiting December 2011. Dr Barnes (hosted by Stuart Ryder) will be working on understanding star formation within the Milky Way, using infrared observations of known cold, dense gas clumps identified through the CHaMP survey.

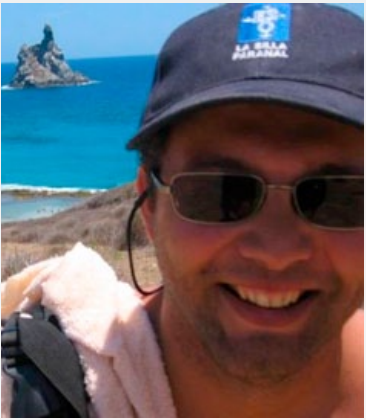

**Dr Claudio Melo**, (ESO), visiting December 2011. Dr Melo (hosted by Gayandhi De Silva) will be working with AAO staff on understanding the origins and formation processes of stellar clusters within the Milky Way.

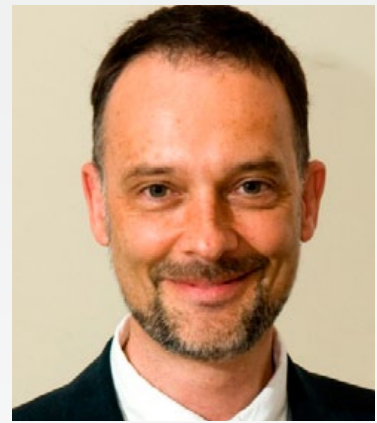

**Professor John Peacock**, (University of Edinburgh), visiting April 2011. Prof. Peacock (hosted by Andrew Hopkins) will be working on cosmological measurements from the GAMA survey.

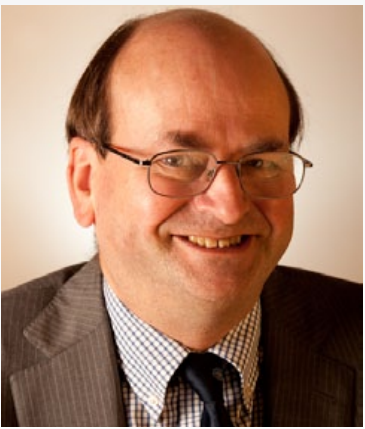

**Professor Jeremy Mould**, (Swinburne University), visiting March-April 2011. Prof. Mould (hosted by Matthew Colless) will be working on mapping the dark matter distribution using the 6dFGS.

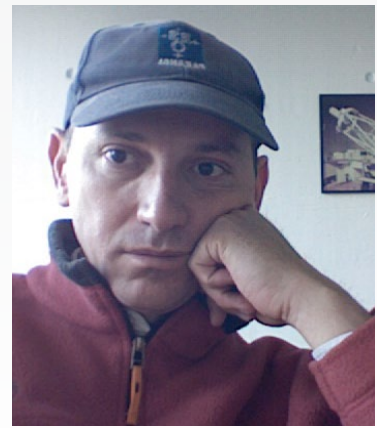

**Dr Jason Spyromilio**, (ESO), visiting January-July 2011. Dr Spyromilio (hosted by Chris Lidman) will be working on measurements of nearby and distant supernovae, as well as various instrumentation projects.

# Epping News

Sarah Brough (AAO)

We have 3 new members of the astronomy group, one ARC-funded Super-Science Fellow **Dr. Amanda Bauer**, AAO Research Fellow **Dr. Maritza Lara-Lopez** and a joint AAO/Macquarie University research Fellow **Dr Angel Lopez-Sanchez.**

**Dr Amanda Bauer** joined the AAO as a Super Science Fellow to work with the GAMA collaboration in November 2010. Amanda grew up in the midwestern city of Cincinnati, Ohio, USA and stayed close to home for her undergraduate degree in Physics at the University of Cincinnati. She made the unexpected move to Austin, Texas after graduation where she completed her PhD in Astrophysics in August 2008 at the University of Texas at Austin. Amanda accepted her first post-doc at the University of Nottingham in England to work with a group of astronomers who are very interested in studying the properties of distant galaxies. Her research has largely focused on galaxies in the distant universe, when they were in their infant and toddler stages of life. GAMA provides a wonderful data set of galaxies in the relatively nearby universe, which she plans to use to study the ways in which galaxies grow up and age over time. Amanda's work here will focus on understanding what physically drives the initiation and termination of star formation inside galaxies in our Universe, and how these processes depend on the masses and shapes of the galaxies, and the neighbourhoods in which they live.

**Dr. Maritza Lara-Lopez** joined the AAO as an AAO Research Fellow to work with the GAMA and Galaxy Genome collaborations in November 2010. Maritza has also been awarded a Super Science Fellowship and will start that in July 2011. Maritza received her PhD in astrophysics from the Instituto de Astrofisica de Canarias (IAC), Tenerife, Spain, in September 2010. The subject of her dissertation is: "Evolution of fundamental parameters of emission-line galaxies up to a redshift 0.4", under the supervision of Dr. Jordi Cepa and Dr. Angel Bongiovanni. During Maritza's PhD thesis she worked on the evolution of the star formation rate (SFR), metallicity, and mass, as well as its relationships, specially the mass-metallicity relation (M-Z). Additionally, she has participated in papers about the evolution of the Tully-Fisher relation and X-ray Studies of AGN. As a new member of the AAO and of the GAMA team, Maritza's research will be focus on the evolution and environmental dependence of the M-Z and Mass-SFR (and Specific SFR) relations in clusters and galaxy pairs using GAMA data.

**Dr Angel Lopez-Sanchez** joined the AAO as a joint AAO/Macquarie University research fellow in January 2011. Angel is originally from Córdoba (Spain) and got his PhD Thesis in 2006 at the Instituto de Astrofísica de Canarias / La Laguna University (Tenerife, Spain). He worked at the CSIRO Astronomy and Space Science (Australia Telescope National Facility) between 2007 and 2010, obtaining radio and optical data of galaxies of the Local Volume. He has a lot of experience supporting students and giving lectures and talks about Astronomy and considers outreach to be very important. Angel is also an amateur astronomer and enjoys observing the sky with his eyes, binoculars or small telescopes and taking astronomical pictures using his own equipment. Angel's astrophysical research is focused in the analysis of star formation phenomena in galaxies of the local Universe, especially in dwarf starbursts and spiral galaxies. He takes a multi-wavelength approach and hence combines ultraviolet, optical, infrared and radio data to characterise the physical and chemical properties of galaxies and get clues about their nature and evolution. The main project Angel will be working on during his joint AAO/Macquarie University position is to perform a detailed analysis of the properties (stellar and gas masses, star-formation activity, dust content and metallicity) of a sample of ~80 gas-rich galaxies of the Local Volume (within 10 Mpc). This analysis is needed to get a better understanding of the physical processes than govern the star-formation efficiency of galaxies and to constrain models of galaxy evolution.

The instrumentation group sees the start of a new Group Manager **Dr Pascal Xavier.** Pascal joined the AAO in August 2010. He was working as a Senior Project Manager at GHD, an engineering consulting company (much like SKM), prior to this. A mechanical engineer by training, he went on to specialize in polymer fibre composites obtaining his PhD in 1994 on the mathematical modeling of laminated fibre composites. He worked for about 9 years in the defence R&D industry including an overseas attachment to the Pennsylvania State University, USA for 3 years as a post doc. There he was significantly involved in the design and development of a nanotechnology-based inertial navigation sensor for which a patent was granted in 2001. He worked in an US semiconductor company for 3 years prior to moving to Australia. He obtained an MBA on Intellectual Property and Commercialisation in 2003. In Australia, he worked as a project manager in the mining and infrastructure industry for about 5 years working in such remote areas as Port Hedland, WA, Mount Isa, QLD and Bega, NSW. Pascal is happy to return to the all familiar R&D environment at the AAO, where he is contributing to project and resource management.

We also have two new instrument scientists joining us, **Dr Simon Ellis and Dr Jessica Zheng.**

**Dr Simon Ellis** is a returnee to the AAO, starting this time in October 2010. Simon got his PhD from the University of Birmingham, UK, in 2003 for his thesis on the evolution of high redshift clusters of galaxies. He then moved to the AAO, where he worked for 5 years, first as the Research Fellow, and then as part of the instrumentation group. During this time Simon began working on OH suppression, as well as continuing his interest in galaxy evolution. From 2008 to 2010 Simon worked at the University of Sydney as a Research Fellow, with Joss Bland-Hawthorn. He is now part of the Instrument Science group at the AAO again, where he am currently involved in the design of GNOSIS, which is to be commissioned in March and May 2011. GNOSIS will be the first instrument to employ OH suppression with fibre Bragg gratings, and will deliver an extremely dark near-infrared background. Simon looks forward to pursuing the resulting science and pushing astronomical instrumentation further as part of the AAO.

**Dr Jessica Zheng** joined the AAO as an instrument scientist in September 2010. She obtained her Ph.D on Photonic RF signal processing at 2004 from Western Australia Center of Excellence for MicroPhotonic Systems, Edith Cowan University. Jessica has had more than 10 years' academic and industry work experience in different areas, covering photonic RF signal processing, optical precision instrument design, laser 3D scanner, fibre laser, optical switch by using liquid crystal technology, adaptive optics and imaging system design. She is excited about working in a new research area and will participate in the development of new photonics technologies and concepts for astronomical instrumentation for various telescopes. In the mean time, Jessica will provide instrument development support for AAO's ongoing instrumentation projects.

**Julia Tims** is joining us as a new AAO project manager starting early January 2011. Julia has 10 years of experience in the aerospace and construction areas. She is also bringing technical expertise in instrument testing.

Sadly people have also left the AAO. We wish them all the best in their new endeavours! Testing the cosmological model using the WiggleZ Dark Energy Survey.

AAOmega Instrument Scientist **Dr Rob Sharp** left the AAO in late September after 6.5 years here. He has moved to the Australian National University's Research School of Astronomy and Astrophysics, to work on instrumentation for the Giant Magellan Telescope, among other projects. Rob will retain an affiliation with the AAO as one of our Honorary Associates, and he will continue to actively participate in his many research projects with AAO staff and as well as his own frequent observing programs at the AAT.

**Dr David Floyd** has been an AAO staff member since early 2007, when he was appointed as one of the first Magellan Fellows. He has spent the past 15 months at the University of Melbourne working with Prof. Rachel Webster and colleagues on quasar research. David has been able to extend his stay at the University of Melbourne for the short term.

Project Manager **Jana Mladenoff** has taken a higher profile job in the construction industry and will leave the AAO at the end of January 2011.

## Summer Students

The AAO runs a twice-yearly fellowship programme to enable undergraduate students to gain 10-12 weeks first-hand experience of astronomical-related research. The current crop of students are from all around the world, some escaping the snow and others enjoying spending their summer learning new skills.

**Florent Bastien** is studying at the French Air Force academy to become an engineer officer. He is in his last year of study and has to do an internship which aims at discovering a domain other than school. As he is very interested in space research having done an internship at the French space centre in French Guiana last year, he finds it a pleasure to work on the liquid atmospheric dispersion correctors for 3 months at the AAO. Because of the dispersion of light by the atmosphere, dispersion correctors have to be set up on telescopes, otherwise the different colours of the star we are looking are dispersed and its image is spread. Usually, the correctors are made with glass prisms but for giant telescopes, the required-size of the prisms makes them difficult to build. That's why, the idea of the project is to replace glass prisms by two liquids. However, the two liquids have to have some specific properties which make them difficult to find. Consequently, I am trying, during my studentship, to find the best combination of liquids.

**Brian Baumgartner** is a graduate of the University of California, San Diego in Physics, with a specialization in Experimental Science. He has worked in oceanography and astrophysics labs in the United States, as well as for the U.S. Department of Defense as a Mechanical Engineer. In his summer project with the AAO, Brian is working with Dr. Anthony Horton in the Instrument Science group of AAO studying focal ratio degradation, its causes, and performance of several fiber optic cone shapes in combating losses due to FRD.

**Elise Hampton** is currently a student at the University of Adelaide, studying a degree in Space Science & Astrophysics. She is about to start her third and possibly final year of her degree before embarking on the path to Honours and a PhD in the field of Astronomy/Astrophysics. She is here working under the supervision of Dr. Sarah Brough on major dry mergers in Brightest Cluster Galaxies and non-central galaxies of similar mass. And enjoying every minute of it!

**Scott Thomas** is a student at the University of Canterbury in New Zealand, just about to start the third year of his BSc, dual majoring in astrophysics and biochemistry. He has come to the AAO to work with Simon O'Toole on the SPADES project (Search for Planets Around Detached Eclipsing Systems). This will involve obtaining lightcurves for several binary stars detected by SuperWASP (Wide Angle Search for Planets) and fitting models to them, allowing the calculation of various pieces of information about the system. These results will be used as the basis for a catalogue to allow SPADES observers to analyse these targets for possible planets. The remainder of his time here will be spent searching for evidence of planets around these and other stars.

**Cheryl Wheeler** is a student at the University of Newcastle, Australia where she has graduated with a Bachelor of Science (Photonics) and is currently pursuing a second degree in earth sciences. Cheryl has previously worked with ANSTO conducting research on a project titled "The Origin of Stretched Exponential Dynamics in Hydrated Proline". As a summer student at AAO, she is working with Dr. Michael Goodwin and the Instrument Science Group on the Starbugs project, researching performance characterisation of the Starbugs for MANIFEST.

## Obituary: Ken Gorham, 1940-2011
Russell Cannon (AAO)

We were very sorry to learn that Ken Gorham died on 15 January after a lengthy battle with cancer, as this issue was going to press. Ken joined the AAO as Personnel Officer in 1977 and was a key member of the administration team until he retired 20 years later. His funeral in Brisbane was attended by more than a dozen former AAO colleagues. Ken is fondly remembered for the dedicated and ever-helpful way in which he carried out his duties, his irrepressible sense of humour and his organisation of staff social events, especially the famous Epping BBQs. His objective was always to assist the rest of us to achieve our objectives, and to keep the AAO running smoothly. ✦AAO

# Letter from Coona

Katrina Harley (AAO)

Firstly, I would like to thank everyone at the AAO for making me feel so welcome. Overall, it hasn't been a dramatic transition moving back to Coona, after being away from home for 8 years. The shopping is still a little disappointing, but the traffic is fantastic.

Unfortunately, we haven't seen much of summer so far, just a lot of rain. Not only disappointing for the Astronomers, but also for many of the farmers in the district who are still struggling to get their winter crops harvested.

I'm sure many of you have already heard of our lunches that have become a regular occurrence - as long as **Raelene Suckley** is here to organise us that is. Many of the staff have taken in turns to cook up a feast, from Chinese to a BBQ. With all of the funds raised going to our Social Club.

2010 was the year for the Christmas party to be held in Coona. It was decided that the best location for the occasion would be the Imperial Hotel, at the recently renovated Gecko Red restaurant. At the end of the meal **Darren Stafford** presented a gift to Operations Manager **Doug Gray** from all of the staff – his own jacket from his office with "Doug's Zumba Class" stitched in pink letters on the back. In all, the evening was a hit, with many making their way to the dance floor - to my horror!

Our mechanical assistant **Glen Murphy** left in early November. **Randal Darko** was employed as a casual to replace Glen. Just recently, Randal was named the successful applicant for the apprenticeship position - congratulations Randal!

Telescope Systems Manager **Bob Dean** is the 2011 Citizen of the Year for the town of Coonabarabran, NSW. Robert received the award at a ceremony on Australia Day from the Australia Day Ambassador for the Warrumbungle Shire, Actress Benita Collings. While all of us at the

AAO are grateful for Bob's expertise it was his contribution to the wider Coonabarabran community (including the local community radio station, Rotary Club and bushfire brigades amongst many others) that was recognised on Australia Day.

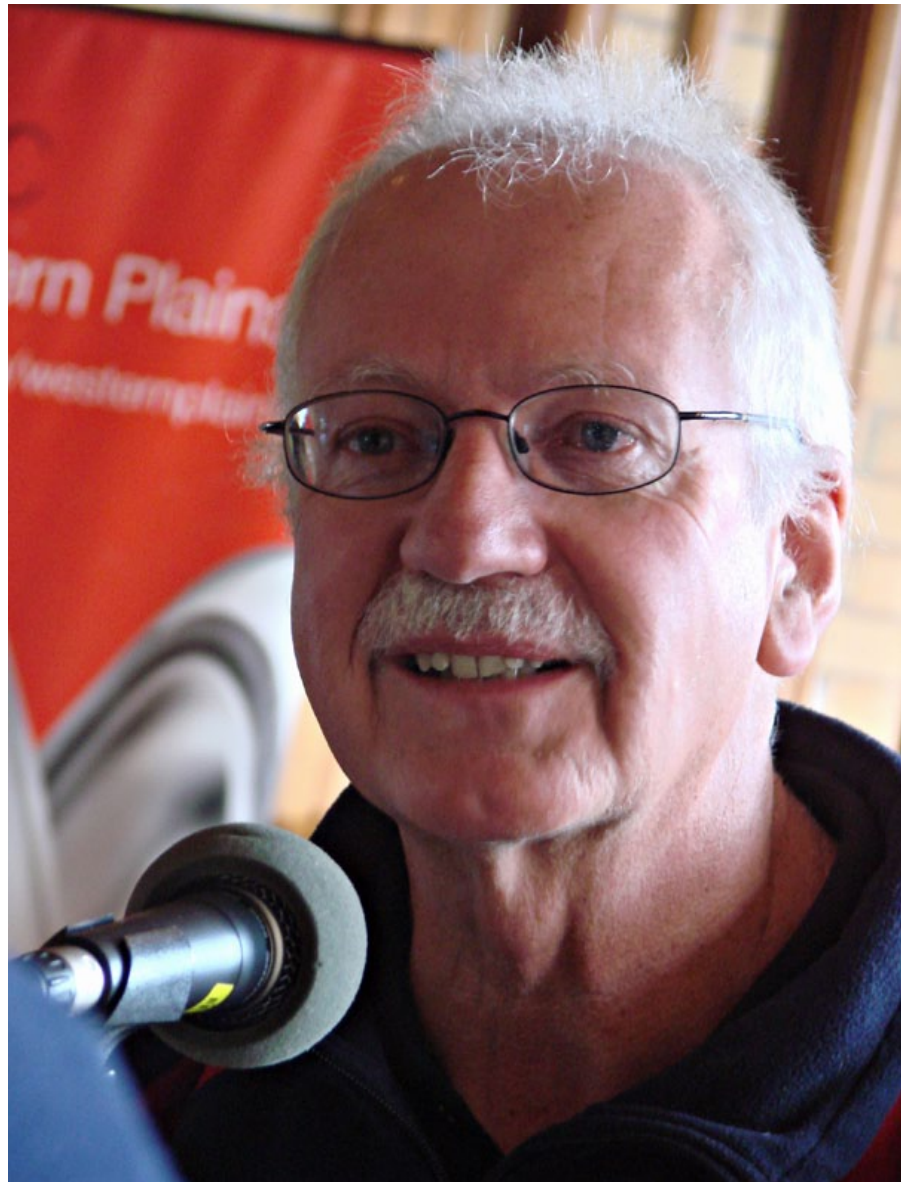

Bob Dean being interviewed by ABC Western Plains in June 2010.
Photo: Justin Huntsdale, ABC Western Plains

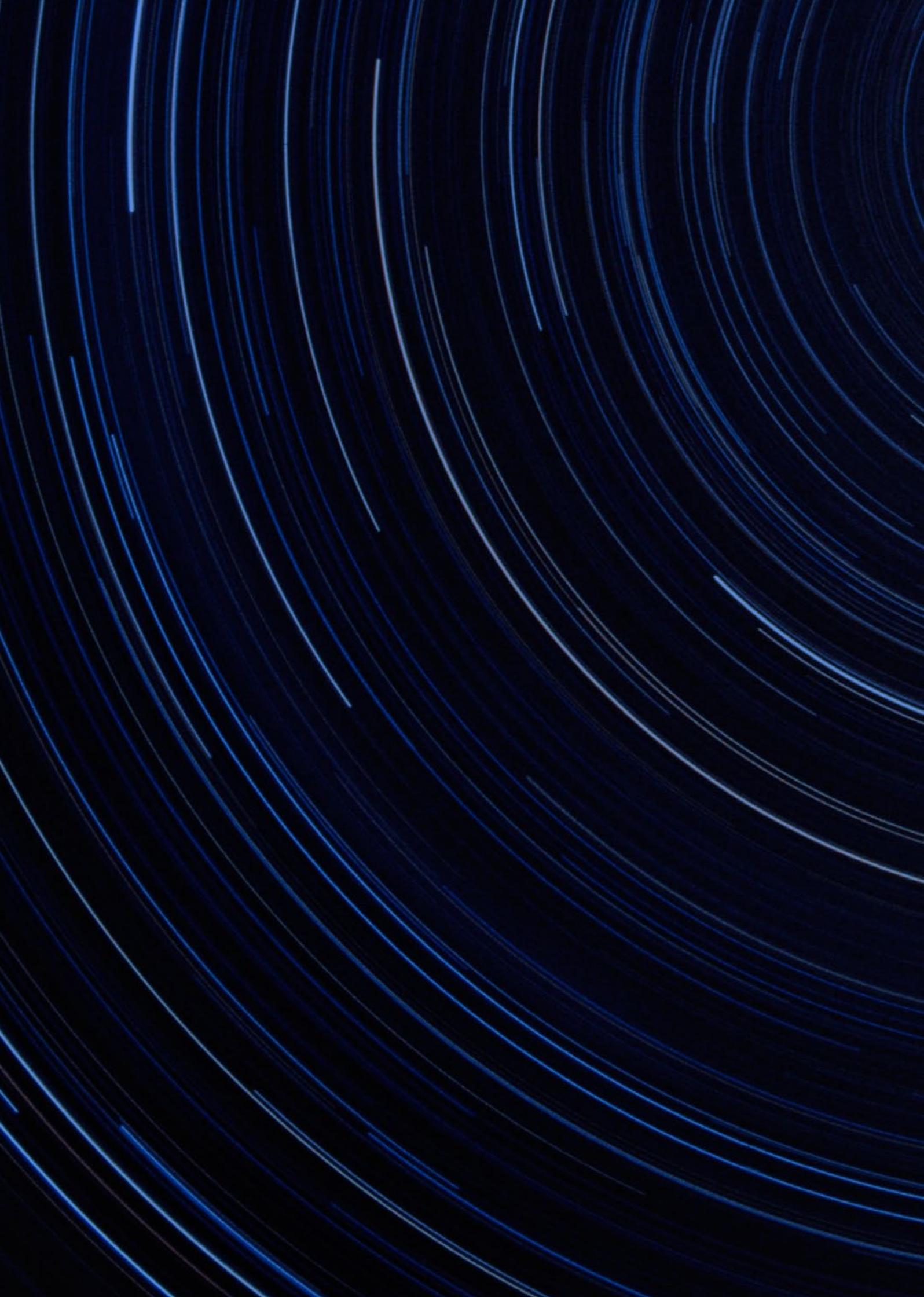

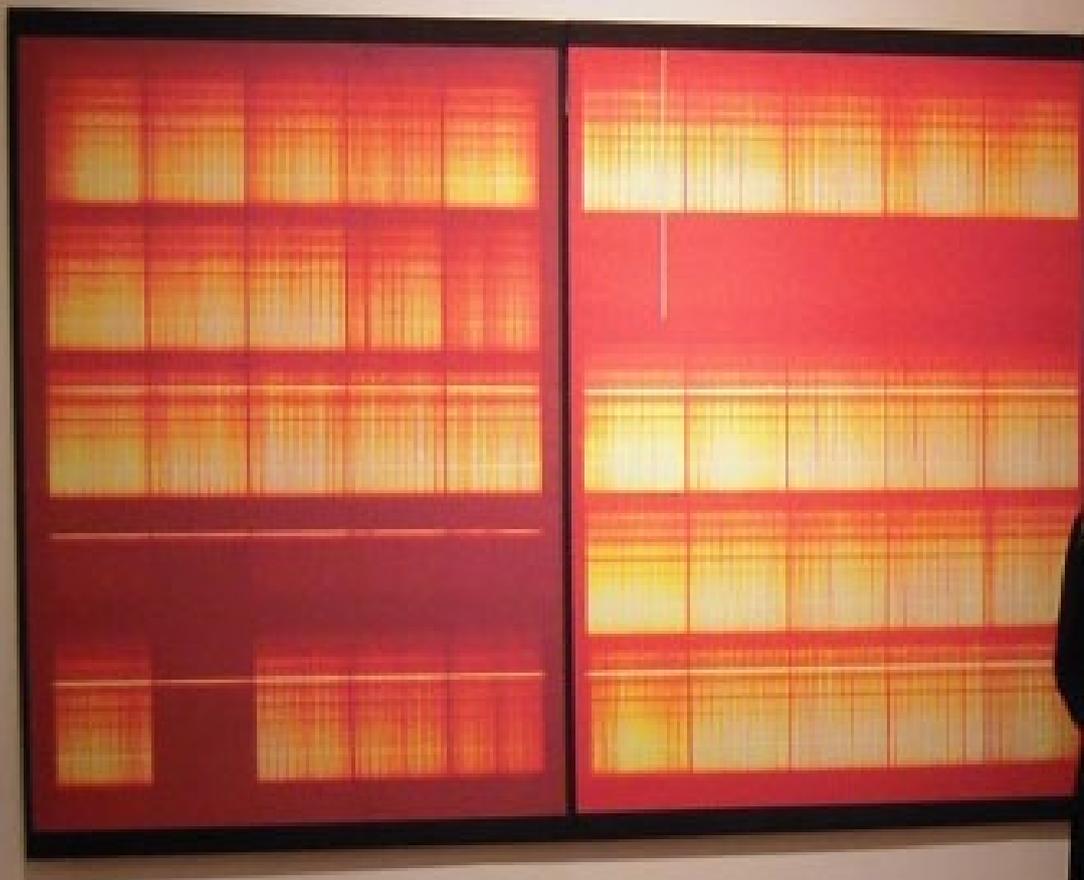

ABOVE: Other uses for astronomy data: Ulrike Kuchner visited the
AAO and University of Sydney in 2009 from Austria. In Austria Ulrike
is studying for her Astronomy Masters degree at Vienna University
and her Fine Arts degree at the University of Applied Arts. Ulrike
graduated from her Fine Art degree in 2010, winning a prize for best
diploma with an edited form of the VLT IFU data she worked on in
Sydney printed on canvas. Congratulations Ulli!

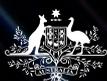

**Australian Government**

**Department of Innovation
Industry, Science and Research**

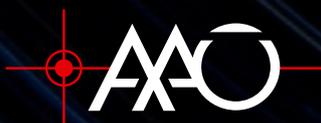